\documentstyle[epsf,rotate,psfig]{mn_plus_bib}

\def\gsim{~\rlap{$>$}{\lower 1.0ex\hbox{$\sim$}}}
\def\HST{{\it HST}}
\def\HIPPARCOS{{\it HIPPARCOS}}

\def\kms{\,\hbox{km}\,\hbox{s}^{-1}}

\def\kpc{\,\hbox{kpc}}
\def\H0{H_0=100 \, h \, {\rm kms^{-1}Mpc^{-1}}}

\def\etal{{et al.\thinspace}}
\def\eg{{e.g.\thinspace}}

\def\ie{{i.e.\thinspace}}

\def\V{{\emph{V}}}

\def\gsim{~\rlap{$>$}{\lower 1.0ex\hbox{$\sim$}}}

\begin{document}

\title{Testing Stellar Population Models with Star Clusters in
the Large Magellanic Cloud}
 
\author[M.A.~Beasley, F.~Hoyle and R.M.~Sharples]
  {Michael A. Beasley$^{1,3}$\thanks{email:
mbeasley@astro.swin.edu.au}, Fiona~Hoyle$^{2,3}$\thanks{email:
hoyle@venus.physics.drexel.edu} and Ray
M. Sharples$^{3}$\thanks{email: R.M.Sharples@dur.ac.uk}\\ 
 $^1$Centre for Astrophysics \& Supercomputing, 
  Swinburne University of Technology, Hawthorn, VIC 3122,
Australia\\   
  $^2$Department of Physics, Drexel University, 3141 Chestnut Street, 
Philadelphia, PA 19104, USA\\
$^3$Department of Physics, University of Durham, Durham DH1 3LE, UK}

\date{Accepted~~~~~~~~~~.   Received~~~~~~~~~~.}

\pagerange{\pageref{firstpage}--\pageref{lastpage}}
\pubyear{2002}

\label{firstpage}

\maketitle

\begin{abstract}
We present high signal-to-noise integrated spectra of 24 
star clusters in the Large Magellanic Cloud (LMC), 
obtained using the FLAIR spectrograph at the 
UK Schmidt telescope. 
The spectra have been placed onto the Lick/IDS system in 
order to test the calibration of Simple 
Stellar Population (SSP) models (Maraston \& Thomas 2000; 
Kurth, Fritz-von Alvensleben \& Fricke 1999).

We have compared the SSP-predicted metallicities 
of the clusters with those from the literature, 
predominantly taken from the Ca-Triplet spectroscopy
of Olszewski \etal (1991).
We find that there is good agreement between 
the metallicities in the range --2.10 $\leq$ [Fe/H] $\leq$ 0. However, 
the Mg$_2$ index (and to a lesser degree
Mg $b$) systematically predict higher metallicities
(up to +0.5 dex higher) than $\langle$Fe$\rangle$. 
Among the possible explanations for this are that 
the LMC clusters possess [$\alpha$/Fe] $>$ 0. 
Metallicities are presented
for eleven LMC clusters which have no previous measurements.

We compare SSP ages for the clusters, derived
from the H$\beta$, H$\gamma$ and H$\delta$ Lick/IDS indices,
with the available literature data, and find 
good agreement for the vast majority.
This includes six old globular clusters in our sample, 
which have ages consistent with their \HST\ CMD ages
and/or integrated colours. 
However, two globular clusters, NGC~1754 and NGC~2005, 
identified as old ($\sim$ 15 Gyr) on the basis
of \HST\ CMDs, have H$\beta$ line-strengths which lead ages
which are too young ($\sim$ 8 and $\sim$ 6 Gyr
respectively). 
These findings are inconsistent with their CMD-derived
values at the 3$\sigma$ level.
Comparison between the horizontal branch morphology
and the Balmer line-strengths of these clusters
suggests that the presence of blue horizontal branch 
stars has increased their Balmer indices
by up to $\sim$ 1.0 \AA.

We conclude that the Lick/IDS indices, used in conjunction 
with contemporary SSP models, are able to reproduce the ages and 
metallicities of the LMC clusters reassuringly well.
The required extrapolations of the fitting-functions
and stellar libraries in the models to younger ages 
and low metallicities do not lead to serious systematic errors.
However, due to the significant contribution
of horizontal branch stars to Balmer
indices, SSP model ages derived for metal-poor 
globular clusters are ambiguous 
without {\it a priori} knowledge of 
horizontal branch morphology. 
\end{abstract}


\section{Introduction}
\label{Introduction}

One of the most powerful tools available to 
observers of stellar populations is the colour-magnitude
diagram (CMD). Whilst there still remain numerous
uncertainties in stellar evolution theory (e.g. \citeANP{Castellani01}
2001), the existence of accurate paralaxes such as those 
from \HIPPARCOS, used in conjunction 
with contemporary model isochrones can now constrain the ages of 
Galactic globular clusters (GCs) to within $\sim$ 20\% 
(e.g. \citeANP{Carretta00} 2000).

Furthermore, \HST\ has allowed us to obtain detailed information 
for GCs and field stars in external galaxies such as the 
LMC (e.g. \citeANP{Holtzman97} 1997; 
\citeANP{Olsen98} 1998; \citeANP{Johnson99} 1999), 
the Andromeda galaxy (e.g. \citeANP{Holland96} 1996; 
\citeANP{FusiPecci96} 1996)
and even the nearest large elliptical Centaurus A (e.g.
\citeANP{Soria96} 1996; \citeANP{Harris98} 1998; 
\citeANP{GHarris99} 1999; \citeANP{Marleau00} 2000).
    
However, such observations are challenging, and beyond 
several Mpc exceed the capabilities of current
instrumentation. Even with \HST, a combination of crowding 
and the intrinsic faintness of single stars limits the 
applicability of such an approach. Therefore, 
in order to probe the properties of distant stellar
systems, we must rely upon studies of integrated light.

Integrated spectroscopy and photometry  
require comparisons with stellar
population models, and are affected by a degeneracy
between age and metallicity 
(\citeANP{Faber72} 1972; \citeANP{OConnell76} 1976). 
Spectroscopic indices have been shown to hold potential, and
much work in the past decade has lead to age-metallicity
diagnostics for integrated spectra (\eg Gonz$\acute{a}$lez~1993;
\nocite{Gonzalez93}\citeANP{Rose94} 1994; \citeANP{Worthey94} 1994; 
\citeANP{Borges95} 1995; \citeANP{Idiart95} 1995;
\citeANP{WO97} 1997; \citeANP{Vaz99} 1999).
These methods have subsequently been used by many workers
to derive ages and metallicities for galaxies 
(\eg \citeANP{Davies93} 1993; Gonz$\acute{a}$lez~1993;
\citeANP{Fisher95} 1995; \citeANP{Harald98} 1998; 
\citeANP{Vazdekis01a} 2001) and
extragalactic globular clusters (\eg \citeANP{Cohen98} 1998; 
\citeANP{KisslerPatig98} 1998; \citeANP{Beasley00} 2000; 
\citeANP{Forbes01b} 2001).

However, the reliability of such integrated techniques has yet
to be demonstrated: they must be tested
against stellar systems with independently derived age
and metallicity estimates such as Galactic GCs.
Addressing this issue, \citeANP{Gibson99} (1999) derived a 
'spectroscopic' age for 47 Tucanae in the H$\gamma_{\rm HR}$ --
Fe4668 and H$\gamma_{\rm HR}$ -- CaI$_{\rm HR}$ planes of the
\citeANP{Worthey94} (1994; henceforth W94) simple
stellar population (SSP) models. These authors found
that 47 Tuc's integrated spectrum fell below the oldest (17 Gyr)
isochrones of the W94 models at the 4$\sigma$ level, yielding
an extrapolated age in excess of 20 Gyr. By comparison, the
CMD-derived age of 47 Tuc is 14 $\pm$ 1 Gyr \cite{Richer96}.
On the other hand, \citeANP{Maraston00} (2000) derived 
an age of 15 Gyrs for this cluster using the combination H$\beta$
and Fe5335, in good agreement with the CMD of \citeANP{Richer96} (1996).
\citeANP{Vazdekis01} (2001) and \citeANP{Schiavon02} (2002) have 
recently addressed these issues and conclude that the 
inclusion of atomic diffusion and non-solar abundance ratios 
are important. Moreover, \citeANP{Schiavon02} (2002) found it
necessary to include both AGB stars and 
adjust the metallicity of 47 Tuc by --0.05 dex to fit their 
SSP models to the integrated spectrum of this cluster.

Whilst these developments are extremely promising,
the full calibration of SSP models has yet to 
be comprehensively tested.
The Galactic GC 47 Tuc represents 
a single age and single metallicity 
in the large parameter space of contemporary 
SSP models. In view of the adjustments employed
Schiavon \etal 2001 in order to reproduce
the integrated spectrum of just this cluster, 
begs the question of how well can these models 
be applied to more distant, less well-known stellar 
systems?
Furthermore, 47 Tuc is an idealised case of an old, 
relatively metal-rich stellar system
which has a 'red clump' for its horizontal branch (HB).
In this case, the strength of the Balmer lines are 
thought to be primarily a function of 
the temperature of the main sequence turn-off (and hence age). 
At lower metallicities\footnote{but see \citeANP{Rich97} (1997)
for two examples of relatively metal-rich Galactic GCs with 
blue HBs.} GCs develop blue HBs which are 
expected to contribute a significant component to 
Balmer lines (\eg \citeANP{Buzzoni89}
1989; \citeANP{deFreitasPacheco95} 1995; \citeANP{Lee00a} 2000; 
\citeANP{Maraston00} 2000). 

On the observational side, obtaining
integrated spectra of Galactic GCs may hide other
uncertainties. Owing to the large angular
size of Galactic GCs on the sky (the half-light radius of
47 Tuc is $\sim$ 2.8 $\arcmin$) a spectroscopic 
aperture must be synthesised by physically scanning
a slit across the cluster. In so doing, the observer 
may unwittingly include foreground stars in the integrated
spectrum, in addition to increasing the liklihood of stochastic
contributions from bright stars. Clearly, alternative 
laboratories are desirable to test SSP models.

The Large Magellanic Cloud (LMC) presents an 
ideal target for such tests. 
Many of its $\sim$ 2000 star clusters \cite{Olszewski96}
have been independently age-dated and their metallicities
determined, whilst its proximity ($\sim$ 53 \kpc) 
means that acquiring high S/N, integrated spectra 
of the clusters is relatively straightforward
(\eg \citeANP{Rabin82} 1982).

In this paper, we present high S/N integrated
spectra for 24 star clusters in the LMC, covering
a wide range in age (0.5 -- 17 Gyr) and metallicity 
(--2.1 $\leq$ [Fe/H] $\leq$ 0).
We have placed the clusters onto the Lick/IDS
system and measured their line-strength 
indices in order to test contemporary stellar 
population models which use the 
Lick/IDS fitting-functions (\citeANP{Gorgas93} 1993; 
\citeANP{Wortheyetal94} 1994;
\citeANP{WO97} 1997).

The plan of this paper is as follows: 
In Section~\ref{sec:SampleSelection}, 
we describe our sample selection and the
observations performed for this present study. 
In Section~\ref{sec:DataReduction} we describe
the reduction steps required for our fibre spectra.
We discuss the spectroscopic system and the stellar population models 
we use in this study in Section~\ref{sec:TheSpectroscopicSystem}.
In Section~\ref{sec:TheAgesandMetallicitiesoftheLMCClusters},
we derive ages and metallicities for the LMC 
clusters using the SSP models, which we then compare to 
literature values.
Finally, we present our conclusions and a summary
in Section~\ref{sec:SummaryandConclusions}.

\section{Sample Selection and Observations}
\label{sec:SampleSelection}

\subsection{Star Clusters in the LMC}
\label{subsec:StarClustersintheLMC}

\citeANP{SWB80}~(1980; henceforth SWB) devised a 
one-dimensional classification 
scheme for LMC star clusters using integrated Gunn-Thuan 
photometry, and showed that the clusters primarily form an age sequence
in the {\it Q(ugr)-Q(vgr)} diagram, with metallicity becoming 
increasingly important for the oldest clusters.
SWB assigned 'SWB-types' I--VII to the clusters, with I
representing young, blue clusters through to VII, 
old and metal-poor clusters---essentially analogues
of Milky Way GCs.

Subsequently, \citeANP{FrenkandFall82}~(1982) showed that the 
same sequence was apparent in the 'equivalent' \emph{(U-B}) vs 
\emph{(B-V)} diagram, which they termed 'E-SWB'. 
\citeANP{Elson85} (1985) and \citeANP{ElsonandFall88} (1988)
determined an age calibration for these SWB types by using
literature age estimates, determined from either 
CMDs, integrated spectra, or from the extent of their 
asymptotic giant branches (\eg \citeANP{Mould82} (1982).
In this paper, we employ SWB-types to 
segregate the LMC clusters into age/metallicity groups
for our analysis, 
using the revised age calibration given in \citeANP{BCD92} (1992).
\footnote{As part of this classification, \citeANP{BCD92} (1992)
split the SWB IV class into IVA and IVB, corresponding
to bluer and redder colours respectively.}

Our spectroscopic sample was selected on the basis
of the availability of independent age and metallicity
estimates for the clusters, SWB type and concentration. 
This final criterion was important so that we were able to 
match the core radii of the clusters to the size of the 
6$\arcsec$ fibres of the FLAIR spectrograph (see below), thereby 
minimising background contamination. 
We took particular care to include 
SWB VII (globular) clusters in our sample, 
which have \HST\ CMDs.
Integrated colours, magnitudes and positions for our cluster
sample were drawn from the catalogue of 
\citeANP{Bica99}~(1999), with SWB-types obtained from the
earlier catalogue of \citeANP{Bica96}~(1996).

Our total spectroscopic sample comprised of SWB I--VII
clusters. However, for the purposes of this analysis, 
only the clusters of type IVA ($\sim$ 0.4 Gyr) and older
will be considered further, since the Lick/IDS 
fitting-functions were only 
calculated for ages of 0.5 Gyr and older (see
\citeANP{Wortheyetal94} 1994). 
We list the 24 clusters in our sample for which 
we have obtained useful spectra in Table~\ref{tab:sample}, 
along with their basic physical parameters.
In addition, we give heliocentric radial velocities for the clusters
obtained from cross-correlation against template stars.
Although the resolution of our spectra is too low
for accurate radial velocities, an approximate velocity
is required in order to shift the spectra to the rest-frame
for the measurement of line-strength indices.
We have achieved S/N ratios of 66 -- 340 \AA$^{-1}$, 
calculated in the 5000\AA\ region of
the spectra (note that for the older clusters  
transmitted flux at the H$\gamma$ feature is typically a 
factor of $\sim$ 2 lower than at 5000\AA, 
\ie a factor of $\sqrt{2}$ lower in S/N.)

\begin{table*}
\begin{center}
\caption[Data for Spectroscopic Sample.]
{Data for our observed star clusters. Columns are : (1) ID
(2) heliocentric velocity, (3) velocity error, (4) right ascension, 
(5) declination, (6) apparent diameter, (7) integrated 
$\V$ magnitude, (8) $U-B$ colour, (9) $B-V$ colour, (10) SWB type, 
(11) signal to noise ratio.
Sources: $^{1}$ this work, 
$^{2}$ from \citeANP{Bica99}~(1999) ,$^{3}$ from \citeANP{Bica96}~(1996)}
\begin{tabular}{lccllcccccl}
\hline \hline
ID & V$_{h}^{1}$ & V$_{err}^{1}$ & RA(2000.0)$^{2}$& DEC(2000.0)$^{2}$& 
D $^{2}$ & $\V^{3}$ & $U-B^{3}$ & $B-V^{3}$& SWB type$^{3}$ & S/N$^{1}$ \\
   & $\kms$  & $\kms$    &           &		  & (arcsec)& (mag) & (mag)     
& (mag) &          &  \AA$^{-1}$     \\
\hline
NGC~1718 & 280 & 71 & 04 52 25 & -67 03 05 & 1.80 & 12.25 & 0.26 & 0.76 & VI&79\\	 
NGC~1751 & 313 & 61 & 04 54 12 & -69 48 25 & 1.60 & 11.73 & 0.00 & 0.00 & V&66\\ 
NGC~1754 & 283 & 52 & 04 54 17 & -70 26 29 & 1.60 & 11.57 & 0.15 & 0.75 & VII&183\\ 
NGC~1786 & 271 & 56 & 04 59 06 & -67 44 42 & 2.00 & 10.88 & 0.10 & 0.74 & VII&340\\ 
NGC~1801 & 289 & 136 & 05 00 34 & -69 36 50 & 2.20 & 12.16 & 0.09 & 0.27 & IVA&155\\ 
NGC~1806 & 246 & 63 & 05 02 11 & -67 59 17 & 2.50 & 11.10 & 0.31 & 0.73 & V & 138\\
NGC~1830 & 324 & 113 & 05 04 39 & -69 20 37 & 1.30 & 12.56 & 0.13 & 0.19 & IVA &66\\
NGC~1835 & 227 & 49 & 05 05 05 & -69 24 14 & 2.30 & 10.17 & 0.13 & 0.73 & VII & 158\\
NGC~1846 & 314 & 66 & 05 07 34 & -67 27 36 & 3.80 & 11.31 & 0.31 & 0.77 & VI & 120\\
NGC~1852 & 303 & 49 & 05 09 23 & -67 46 42 & 1.90 & 12.01 & 0.25 & 0.73 & V & 130\\
NGC~1856 & 265 & 89 & 05 09 29 & -69 07 39 & 2.70 & 10.06 & 0.10 & 0.35 & IVA & 212\\
NGC~1865 & 265 & 100 & 05 12 25 & -68 46 23 & 1.40 & 12.91 & 0.08 & 0.69 & VII & 84\\
NGC~1872 & 273 & 131 & 05 13 11 & -69 18 43 & 1.70 & 11.04 & 0.06 & 0.35 & IVA & 93\\
NGC~1878 & 251 & 94 & 05 12 49 & -70 28 18 & 1.10 & 12.94 & 0.07 & 0.29 & IVA & 157\\
NGC~1898 & 254 & 61 & 05 16 42 & -69 39 22 & 1.60 & 11.86 & 0.08 & 0.76 & VII & 77\\
NGC~1916 & 374 & 89 & 05 18 39 & -69 24 24 & 2.10 & 10.38 & 0.18 & 0.78 & VII & 143\\
NGC~1939 & 296 & 72 & 05 21 25 & -69 56 59 & 1.40 & 11.78 & 0.09 & 0.69 & VII & 145\\
NGC~1978 & 318 & 40 & 05 28 45 & -66 14 10 & 4.00 & 10.70 & 0.25 & 0.78 & VI & 150\\
NGC~1987 & 322 & 87 & 05 27 17 & -70 44 08 & 1.70 & 12.08 & 0.23 & 0.54 & IVB & 121\\
NGC~2005 & 279 & 65 & 05 30 09 & -69 45 08 & 1.60 & 11.57 & 0.20 & 0.73 & VII & 82\\
NGC~2019 & 294 & 74 & 05 31 56 & -70 09 34 & 1.50 & 10.86 & 0.16 & 0.76 & VII &103\\ 
NGC~2107 & 312 & 101 & 05 43 11 & -70 38 26 & 1.70 & 11.51 & 0.13 & 0.38 & IVA &154\\ 
NGC~2108 & 276 & 65 & 05 43 56 & -69 10 50 & 1.80 & 12.32 & 0.22 & 0.58 & IVB & 96\\
SL~250 & 325 & 80 & 05 07 50 & -69 26 06 & 1.00 & 13.15 & 0.21 & 0.59 & IVB & 120\\
\hline
\end{tabular}
\label{tab:sample}
\end{center}
\end{table*}

\begin{figure}
\centering
\centerline{\psfig{file=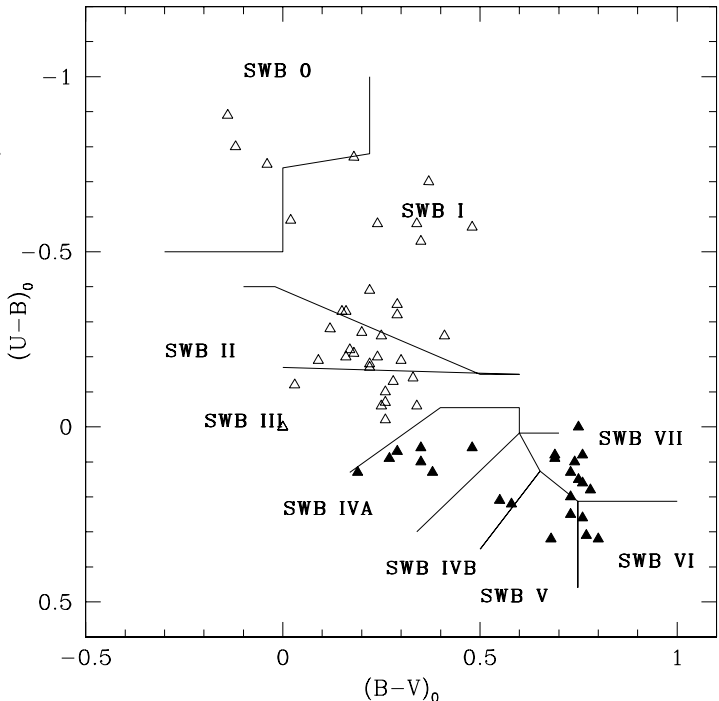,height=10cm}}
\caption{The SWB classifications of LMC star clusters in
the $U-B$,$B-V$ plane. Triangles represent the position of our entire
cluster sample, filled triangles indicate the positions of the
clusters considered in this study. The age calibration for these
SWB types is taken from Bica, Claria \& Dottori (1992), and is given in 
Table~\ref{tab:swb.ages}.}  
\label{fig:swb.sample}
\end{figure}

In Figure \ref{fig:swb.sample} we show the distribution of our
cluster sample in the $U-B$, $B-V$ colour planes. The different
SWB classifications are marked in the figure. The 24 clusters analysed in
this present study are marked as filled triangles, and are all
SWB-type IVA or later. Note that, as one goes to older ages 
(higher SWB types), age and metallicity effects become more
entangled, increasing the likelihood of misclassification of 
the clusters (\eg see Section~\ref{subsec:TheLociofSWB-TypesontheSSPGrids}).

\subsection{The FLAIR System}
\label{subsec:TheFLAIRSystem}

The observations were performed
with the Fibre-Linked Array-Image Reformatter (FLAIR) system
\cite{ParkerandWatson95}, a multi-object fibre spectroscopy 
system at the AAO's 1.2-metre UK Schmidt Telescope (UKST).
The Schmidt photographic plates possess a useful
field area of 40 deg$^{2}$, with a radius of the un-vignetted
field of 2.7 deg. Our candidate clusters were identified
on the Schmidt plates and FLAIR fibres were attached to copies of 
these plates using magnetic buttons. 

\begin{table}
\begin{center}
\centering
\caption[Instrumental Setup]
{The instrumental setup for spectroscopy.}
\begin{tabular}{ll} 
\hline 
Telescope & 1.2-metre UK Schmidt\\
Instrument & FLAIR spectrograph\\
Dates & 3--7 November 1999\\
\hline
Spectral range & 4000--5500 \AA \\
Grating & 600 V\\
Dispersion & 2.62 \AA~pixel$^{-1}$\\
Resolution (FWHM) & $\sim$ 6.7 \AA \\
Detector & EEV CCD02-06 (400 $\times$ 578 pixels)\\
Gain & 1.0 e$^{-}$~ADU$^{-1}$\\
Readout noise & 11 e$^{-}$ (rms)\\
Seeing & 2--3 arcsec \\
\hline
\end{tabular}
\label{tab:setup}
\end{center}
\end{table}

The observations of the star clusters in the LMC were carried out over 
the nights of the 3--7$^{th}$ November 1999.
The general set-up of the FLAIR system for these observations is
given in Table~\ref{tab:setup}.
At the beginning and end of each night, multiple bias-frames
were taken in order to correct for large-scale variations
on the EEV chip. To correct for differences in fibre-to-fibre
response, dome flats and twilight frames were obtained. 
Mercury-Cadmium-Helium arcs were taken for wavelength 
calibration of our final spectra, these were obtained before and
after each target field was observed to check for flexure
or systematic shifts in the spectrograph (this proved unnecessary, 
the FLAIR spectrograph is mounted on the floor and is very stable.)
We opted for the 600V grism used in the first order, with an 
instrumental resolution of $\sim$ 2.6 \AA~pixel$^{-1}$, 
yielding a useful spectral range of 4000--5500 \AA. 
Profile fits to 12 mercury-cadmium-helium arc-lines allowed us to 
determine a full-width half-maximum (FWHM) resolution of the 
spectra of 6.5 $\pm$ 0.2 \AA.

The plate configuration for our two spectroscopic
fields (LMC 1 and LMC 2) are given in Table~\ref{tab:observations}.
The plates cover the same area of sky, but 
represent different fibre configurations.
Eleven {\it different} fibres were assigned to the same clusters
in both fields to provide a measure our repeatability.
This is vital for an accurate assessment of uncertainties, 
since at high S/N ratios, the limiting error in line-strength
indices can stem from night-to-night variations in the
spectroscopic system. A number of dedicated sky fibres were
also assigned for sky-subtraction purposes.

\begin{table}
\begin{center}
\caption[Log of Observations]
{Log of observations for the two plate configurations.}
\begin{tabular}{lccccc} 
\hline 
Field & RA & DEC & Clusters & Sky & Exposure \\
      & (2000) & (2000) & Observed & Fibres & (s) \\
\hline
LMC 1 & 05 23.6 & -69 45  & 39 & 6 & 17100 \\
LMC 2 & 05 23.6 & -69 45  & 39 & 6 & 6300 \\ 
\hline
\end{tabular}
\label{tab:observations}
\end{center}
\end{table}

In addition to our target clusters, we observed a total of 14
standard stars. We obtained spectra for 11 Lick standard stars
covering a range of spectral types and metallicities,  
to calibrate our line-strength indices onto the Lick/IDS system, 
and 3 radial velocity standards. 
We list the observations of our standard stars in
Table~\ref{tab:standards}. For completeness, 
we also include their literature radial velocities and metallicities
where available. Unfortunately, we were unable to 
observe a range of standards covering the full metallicity
range of the LMC clusters in our spectroscopic sample.
However, as shown in $\S$~\ref{subsec:TestingtheLick/IDSCalibration}, 
no gross systematic offsets are introduced into our analysis 
because of this.

\setlength{\tabcolsep}{5pt}
\begin{table}
\begin{center}
\caption[Log of Standard Star Observations]
{Log of observations for Lick/IDS and radial velocity standard stars.
$\dag$ from \textsc{simbad} database. RV = radial velocity standard, 
Lick = Lick standard.}
\begin{tabular}{lllrrl} 
\hline 
ID 	 & Alt. ID	& Spectral  & V$_{r}$ & [Fe/H] & Notes \\
         &              & Type$\dag$& (kms$^{-1}\dag$)&        & \\
\hline
HD~693 	 & HR~33	& F5V		& +14.4 & ... & RV  \\ 
HD~1461	 & HR~72 	& G0V		& --10.7 & --0.33 	& Lick  	\\
HD~4128  & HR~188	& K0III		& +13.0 & ... & RV  \\
HD~4307  & HR~203	& G2V 		& --12.8 & --0.52 & Lick  \\
HD~4628	 & HR~222   	& K2V		& --12.6 & ... 	& Lick  \\
HD~4656	 & HR~224   	& K4IIIb	& +32.3 & --0.07 & Lick  \\
HD~6203	 & HR~296   	& K0III		& +15.3	& --0.48 & Lick  \\
HD~10700 & HR~509   	& G8V		& --16.4 & --0.37	& Lick  \\
HD~14802 & HR~695   	& G2V		& +18.4	& --0.17 & Lick  \\
HD~17491 & HR~832   	& M5III		& --14.0 & ... 	& Lick  \\
HD~22484 & HR~1101  	& F9IV-V	& +27.6	& +0.02	& Lick  \\
HD~22879 & ...	 	& F9V		& +114.2 & --0.85 & Lick  \\
HD~23249 & HR~1136  	& K0IV		& --6.1	& +0.02	& Lick  \\
HD~203638 & HR~8183	& K0III		& +22.0	& ... & RV  \\
\hline
\end{tabular}
\label{tab:standards}
\end{center}
\end{table}

\section{Data Reduction}
\label{sec:DataReduction}

These CCD data were reduced using the \textsc{dofibers}
and \textsc{flair} packages within \textsc{iraf}.
The object frames where all trimmed and bias-corrected
using the multiple bias-frames taken at the time of
the observations. The first-order fibre-to-fibre
response was corrected using the \textsc{triangle}
task. Flat-field and arc-lamp spectra were 
extracted concurrently with the 
object spectra.

For the flat-field corrections, pixel-to-pixel variations 
in the detector were dealt with as well as the differences in 
throughput between the fibres.
A response normalisation function for the fibres was calculated
from the extracted flats, and the flat field spectra were 
normalised by the mean counts of all the fibres, after division by
the response spectra. Because of the differing fibre-to-fibre
response of FLAIR, the spectra were not flux-calibrated, 
but left with the instrumental response characteristic
of the spectroscopic system. As shown in
$\S$~\ref{subsec:TestingtheLick/IDSCalibration} this is 
corrected for in our final line-strength indices.

The spectra were wavelength calibrated using the 
mercury-cadmium-helium arc-lines taken between exposures.
The rms in the fit to the 12 arc features was 
typically $\sim$ 0.1 \AA. The spectra were then 
re-binned onto a linear wavelength scale.
Prior to measuring line-strength indices, 
the spectra were corrected to their rest-frame 
wavelengths using the radial velocities given 
in Table~\ref{tab:sample}.

\begin{figure}
\centering
\centerline{\psfig{file=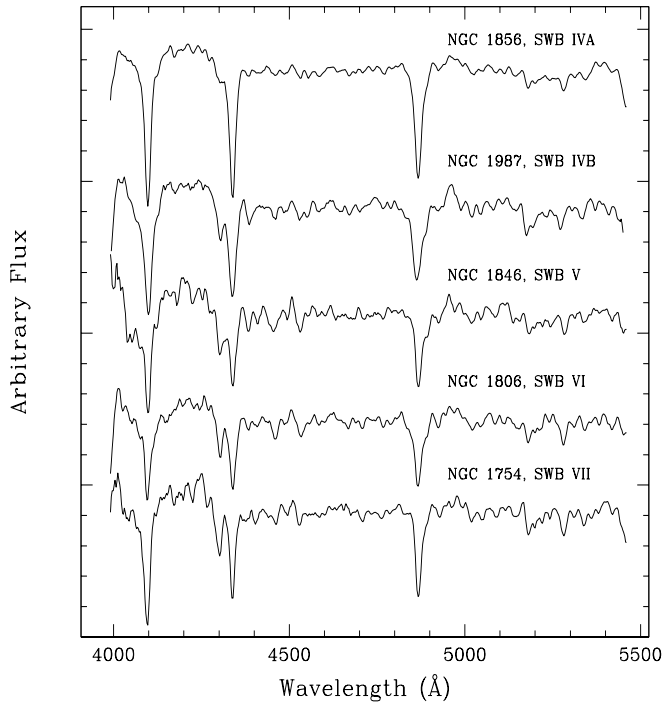,height=10cm,width=9cm}}
\caption{Integrated spectra of the LMC clusters, 
representing each SWB type. The spectra have been smoothed to the
resolution of the Lick/IDS system, and have been
continuum-normalised by division with a low-order polynomial.}
\label{fig:specs}
\end{figure}

Examples of our spectra for SWB-types IVA -- VII
are shown in Figure~\ref{fig:specs}.
A principle feature of the spectra are the weakening
of the Balmer-lines as one goes to later SWB types.
It is also interesting to note the rapid change in the 
spectra between SWB IVA and SWB IVB, characterised
by the developing G-band at 4300 \AA, and 
the increasing prominence of other metallic-lines.

\section{The Spectroscopic System}
\label{sec:TheSpectroscopicSystem}

\subsection{The Lick/IDS Indices}
\label{subsec:TheLick/IDSIndices}

Comprehensive discussions of the Lick/IDS
absorption-line index system, the derived
absorption-line index fitting functions and observations
of stars, galaxies and GCs have been
given in a series of papers by the Lick group
(\citeANP{Burstein84}~1984; \citeANP{Faber85}~1985;
  \citeANP{Burstein86}~1986; \citeANP{Gorgas93}~1993; 
\citeANP{Worthey94}~1994; \citeANP{Trager98}~1998). 

\begin{figure*}
\centering
\centerline{\psfig{file=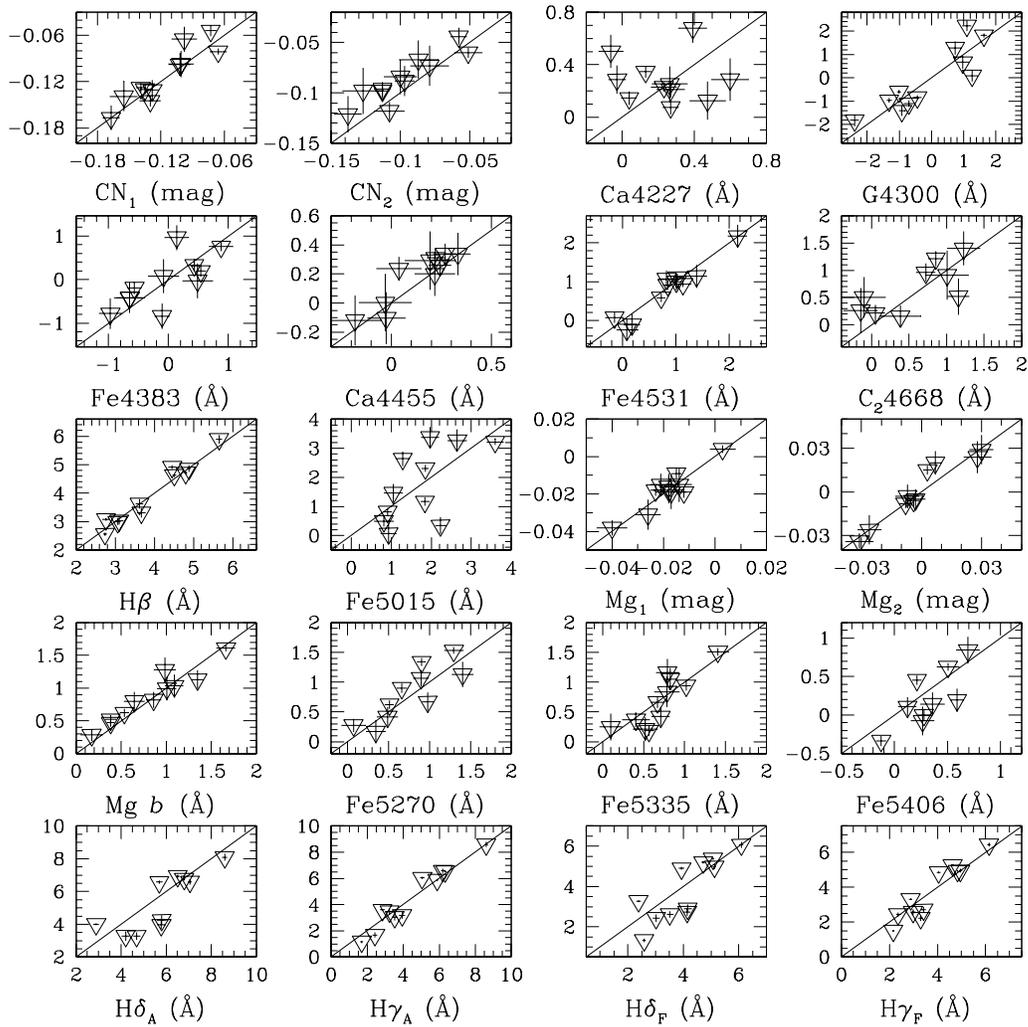,height=15cm}}
\caption{The repeatability of our measurements
of Lick/IDS indices between fields LMC 1 ($x$-axis) and 
LMC 2 ($y$-axis), compared in each case to unit slope (solid line).
The uncertainties on the measurements are derived
purely from Poisson statistics.}
\label{fig:repeats}
\end{figure*}

We have measured 20 Lick/IDS indices, 16
of which are listed in \citeANP{Trager98} (1998), 
supplemented by a further 4 indices defined
in \citeANP{WO97} (1997) which measure
higher-order Balmer lines H$\gamma_{\rm A}$, 
H$\delta_{\rm A}$ (40 \AA\ wide feature bandpass)
and H$\gamma_{\rm F}$, H$\delta_{\rm F}$ 
(a narrower 20 \AA\ wide bandpass).
Due to the high S/N of our data, we derive 
our uncertainty in the measured indices 
from two sources: 
We use Poisson statistics to determine the photon statistical
error in our data,
both in terms of measuring counts in the feature and in the placing
of the continuum bandpass (\eg \citeANP{Rich88} 1988). 
This is added in quadrature to duplicate observations 
of the LMC clusters to assess the
repeatability of our measurements. We show comparisons
between the Lick/IDS indices measured for clusters 
common between fields LMC 1 and LMC 2 in Figure~\ref{fig:repeats}.
To quantify this repeatability, we list the mean offset, standard
deviation on the mean and linear 
correlation co-efficient ({\it r}) for each index in
Table~\ref{tab:repeats}.

\begin{table}
\begin{center}
\caption[Repeatability of the Lick/IDS indices.]
{Repeatability of the Lick/IDS indices between FLAIR fields. }
\begin{tabular}{llr} 
\hline 
Index & LMC 1--LMC 2 & $r$ \\ 
\hline
CN$_1$ (mag) & --0.0054 $\pm$ 0.0043 & 0.917 \\ 
CN$_2$ (mag) & --0.0142 $\pm$ 0.0031 & 0.655 \\  
Ca4227 (\AA) & --0.052 $\pm$ 0.061 & --0.081 \\ 
G4300  (\AA) & --0.036 $\pm$ 0.194 & 0.882 \\     
Fe4383 (\AA) & --0.018 $\pm$ 0.121 & 0.722 \\     
Ca4455  (\AA) & +0.021 $\pm$ 0.023 & 0.960 \\    
Fe4531  (\AA) & +0.052 $\pm$ 0.061 & 0.957 \\      
C$_2$4668  (\AA) & --0.214 $\pm$ 0.183 & 0.666 \\     
H$\beta$  (\AA) & +0.118 $\pm$ 0.067 & 0.978 \\      
Fe5015  (\AA) & --0.005 $\pm$ 0.300 & 0.631 \\  
Mg$_1$ (mag)  & --0.0006 $\pm$ 0.0012 & 0.920 \\
Mg$_2$ (mag)  & --0.0011 $\pm$ 0.0017 & 0.983 \\ 
Mg~$b$  (\AA) & --0.044 $\pm$ 0.034 & 0.958 \\     
Fe5270  (\AA) & --0.139 $\pm$ 0.105 & 0.838 \\    
Fe5335  (\AA) & +0.036 $\pm$ 0.090 & 0.850 \\ 
Fe5406  (\AA) &	+0.183 $\pm$ 0.105 & 0.726 \\     
H$\delta_{\rm A}$ (\AA) &  +0.539 $\pm$ 0.311 & 0.830 \\     
H$\gamma_{\rm A}$ (\AA) & --0.004 $\pm$ 0.167 & 0.942 \\  
H$\delta_{\rm F}$  (\AA)&  +0.265 $\pm$ 0.230 & 0.860 \\     
H$\gamma_{\rm F}$ (\AA) & +0.048 $\pm$ 0.161 & 0.971 \\    
\hline
\end{tabular}
\label{tab:repeats}
\end{center}
\end{table}

Inspection of Figure~\ref{fig:repeats} and 
Table~\ref{tab:repeats} indicates that our repeatability 
for the indices is variable, from excellent 
(\eg H$\beta$, Mg $b$) to very poor (\eg Fe5015). 
However, the majority of the Lick/IDS indices 
do show good agreement.
For N = 11, an $r$
value of 0.8 corresponds to a 3$\sigma$
significant linear correlation \cite{Taylor82}. Therefore, we take
$r \geq$ 0.8 as the threshold for our being able 
to reproduce any given index at high confidence.
 
Our inability to accurately reproduce Ca4227, and to
a lesser extent Fe5015, 
may be due to the narrow nature of their 
bandpass definitions.
For example, the Ca4227 feature is 12.5 \AA~($\sim$5 pixels) 
wide, and its blue-red continuum endpoints bracket only 40 
\AA~($\sim$ 15 pixels). 
As indicated by \citeANP{Tripicco95}~
(1995), Ca4227 is both very sensitive to 
bandpass placement, in addition to
being sensitive to spectral resolution.

\subsection{Calibrating to the Lick/IDS System}
\label{subsec:CalibratingtotheLick/IDSSystem}

The Lick/IDS spectra were all obtained with the 3-metre 
Shane Telescope at the Lick Observatory (hence the Lick system), 
using the Cassegrain Image Dissector Scanner (IDS) spectrograph.
The spectra have a wavelength-dependent resolution 
(FWHM) of 8--10 \AA\ (increasing
at both the blue and red ends)
and cover a spectral  region of 4000--6400 \AA.
The original Lick/IDS spectra are not flux-calibrated, 
but were normalised by division with a quartz-iodide tungsten lamp. 
These two idiosyncrasies of the IDS require that
data from other telescopes be accurately corrected
onto the Lick/IDS system.

\begin{figure*}
\centering
\centerline{\psfig{file=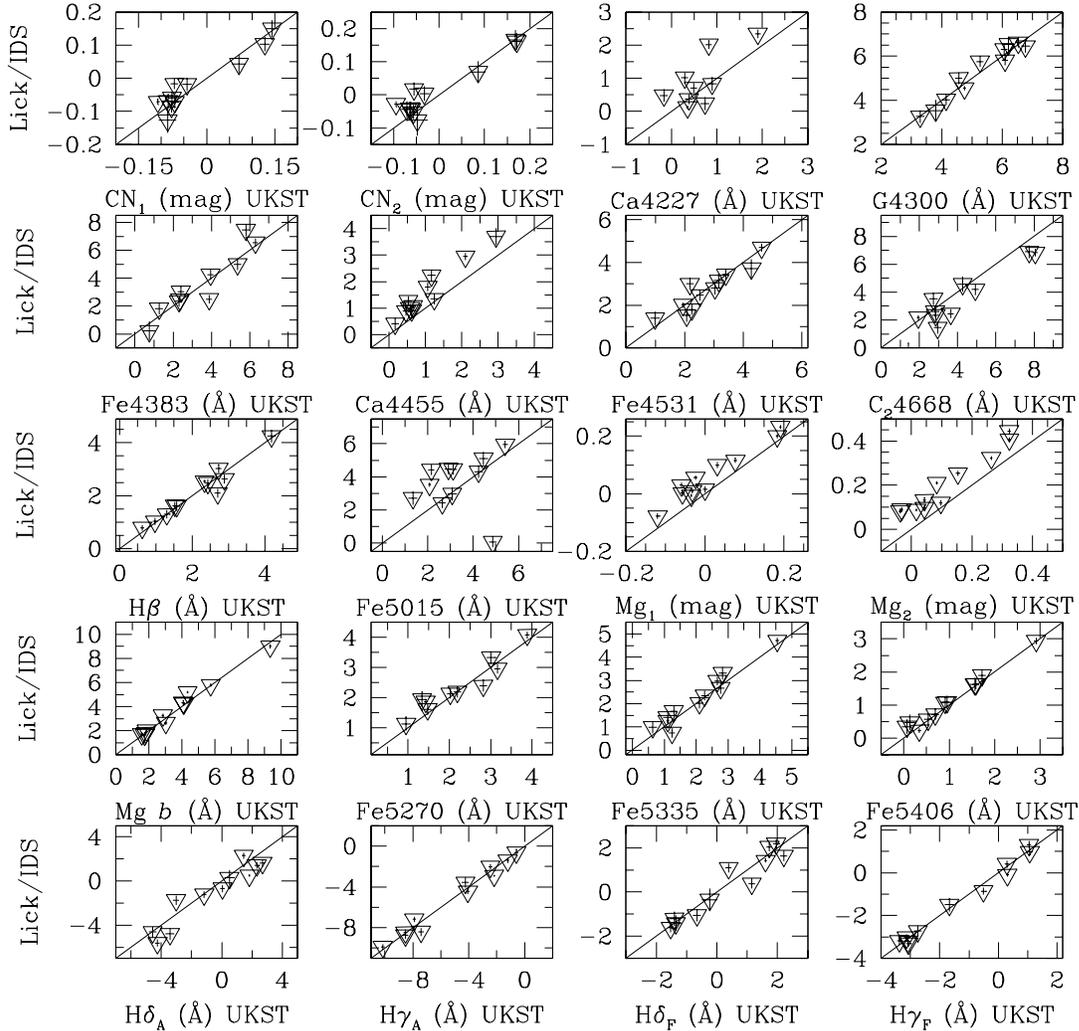,height=15cm}}
\caption{Comparison between the standard star Lick/IDS indices
measured with FLAIR, and the corresponding Lick values. 
Indices are those measured prior to applying the offsets listed
in Table~\ref{tab:offsets}. The solid line indicates unit slope
in each case. Note that the Poisson uncertainties are smaller
than the symbol size in most cases.}
\label{fig:standards}
\end{figure*}

So as to reproduce the resolution characteristics of the IDS
as shown in \citeANP{WO97} (1997), we convolved our 
spectra with a wavelength-dependent Gaussian kernel.
Since the FLAIR spectra have a useful wavelength
range of 4000\AA\ -- 5500\AA, our spectra were 
correspondingly broadened to the Lick/IDS 
resolution of 8.5 \AA\ -- 11.5 \AA.
To remove any remaining systematic offsets between
FLAIR and the Lick/IDS we observed 11 Lick standard
stars (Table~\ref{tab:standards}). For these
stars we measured all the Lick/IDS indices 
within our useful wavelength range, and compared them to the 
tabulated values to obtain additive offsets to apply
to our data. We compare our measured Lick indices for our 
FLAIR standard stars with the corresponding Lick values in 
Figure~\ref{fig:standards}. In Table~\ref{tab:offsets}
we then list the offsets required to achieve zero offset, and
thereby calibrate our data onto the Lick/IDS system.
 
\begin{table}
\begin{center}
\caption[Lick/IDS Offsets]
{Systematic offsets in indices between Lick/IDS and UKST.
Uncertainties are the standard deviation on the mean of each
index.}
\begin{tabular}{ll} 
\hline 
Index & Lick--UKST\\
\hline
CN$_1$ (mag) & --0.0212 $\pm$ 0.004  \\
CN$_2$ (mag) & +0.0110 $\pm$ 0.005 \\
Ca4227 (\AA) & +0.30 $\pm$ 0.17 \\
G4300  (\AA) & +0.06 $\pm$ 0.19 \\
Fe4383 (\AA) & +0.36 $\pm$ 0.12 \\
Ca4455  (\AA) & +0.30 $\pm$ 0.10 \\
Fe4531  (\AA) & +0.40 $\pm$ 0.06 \\
C$_2$4668  (\AA) & --0.10 $\pm$ 0.18 \\
H$\beta$  (\AA) & +0.05 $\pm$ 0.14 \\
Fe5015  (\AA) & +0.81 $\pm$ 0.32  \\
Mg$_1$ (mag)  & +0.041 $\pm$ 0.001\\  
Mg$_2$ (mag)  & +0.065 $\pm$ 0.003 \\ 
Mg~$b$  (\AA) & +0.60 $\pm$ 0.07 \\ 
Fe5270  (\AA) & +0.19 $\pm$ 0.11\\  
Fe5335  (\AA) & +0.33 $\pm$ 0.10\\  
Fe5406  (\AA) & +0.35 $\pm$ 0.11\\  
H$\delta_A$ (\AA) & +0.48 $\pm$ 0.31\\ 
H$\gamma_A$ (\AA) & --0.24 $\pm$ 0.17 \\
H$\delta_F$  (\AA) & +0.30 $\pm$ 0.26 \\
H$\gamma_F$ (\AA) & --0.05 $\pm$ 0.16 \\
\hline
\end{tabular}
\label{tab:offsets}
\end{center}
\end{table}

A number of the offsets between the Lick/IDS measurements 
and those obtained here are substantial, in particular the
CN$_1$, CN$_2$, Mg$_1$ and Mg$_2$ indices. All of these
indices possess wide side-bands, implying that the  
origin of the offsets are continuum slope differences
due to the fact the FLAIR spectra are not flux-calibrated.
As shown in $\S$~\ref{subsec:TestingtheLick/IDSCalibration}, the additive 
offsets given in Table~\ref{tab:offsets} calibrate these indices
to the Lick/IDS system.
The final corrected Lick/IDS indices for the 
LMC star clusters and their associated uncertainties
are given in Table~A1.

\subsection{The SSP Models}
\label{subsec:TheSSPModels}

The models we use in this study are those 
of \citeANP{Maraston00} (2000), \citeANP{Kurth99} 
(1999; henceforth KFF99) and W94. 
By definition, SSP models assume a single
burst of star formation, which occurs 
in the first time-step of the model, subsequently
followed by 'passive' evolution of the stellar population.
The above models which assume a Salpeter IMF  
are adopted, although the exact choice of IMF 
is at best a second-order
effect when compared to uncertainties in the
fitting-functions and mass-loss parameter
(\eg \citeANP{Lee00a} 2000; \citeANP{Maraston01a} 2001).

The evolutionary synthesis models 
of \citeANP{Maraston00} (2000) deal
with post-main sequence stellar evolution using the 
Fuel Consumption Theorem \cite{Renzini86}.
The 'thermal-pulsing asymptotic giant branch'
phase is calibrated upon observations of 
Magellanic Cloud star clusters.
The \citeANP{Maraston00} (2000) models adopt 
classical non-overshooting stellar tracks \cite{Cassisi98}.
These models span a metallicity
range of --2.25 $\leq$ [Fe/H] $\leq$ +0.67, 
and an age range from 30 Myr to 15 Gyr.
In deference to the
previous discussions on the limitations of the 
Lick/IDS fitting functions, we only consider 
models with ages $\geq$ 0.5 Gyr.

The KFF99 models employ isochrone synthesis, 
using the stellar tracks from the Padova group (\citeANP{Fagotto94}
1994, and references therein) which incorporate
convective overshoot. A Monte-Carlo method is employed by these
authors to avoid the problems of interpolating between 
isochrones.
The KFF99 models cover an age-metallicity parameter
space of 0.5 $\leq$ t $\leq$ 16 Gyr, and 
--2.3 $\leq$ [Fe/H] $\leq$ +0.4. Unfortunately, 
these models do not predict line-strength indices
for the H$\gamma$ and H$\delta$ indices.

The original population synthesis
models of W94 cover an age-metallicity parameter space of
1.5  $\leq \tau \leq$ 17 Gyr
and --0.5 $\leq$ [Fe/H] $\leq$ 0.5.
To encompass old, metal-poor populations, these models 
were subsequently extended to bracket --2.0 $\leq$ [Fe/H] $\leq$ --0.5 
for ages 8  $\leq \tau \leq$ 17 Gyr,  calibrated using
Galactic GCs.
The W94 models do not cover the young, metal-poor 
characteristics of the majority of the LMC clusters, 
making them of value only for the old GCs in this present study.
However, since the W94 models predict line-strength
indices for the full range of Lick/IDS fitting functions, 
they are invaluable for checking the calibration of 
these indices.

\subsection{Testing the Lick/IDS Calibration}
\label{subsec:TestingtheLick/IDSCalibration}

Prior to obtaining age and metallicity estimates
for the LMC clusters using the SSP models, 
it is important to ensure that we have been 
able to correct our data onto the Lick/IDS system.
To this end, we compare different Lick/IDS indices of the
clusters which measure similar chemical species (or are at least influenced
by these same elements \eg~\citeANP{Tripicco95} 1995).
By plotting these indices against each other onto SSP grids, 
which are effectively degenerate in age and metallicity, 
we can look for evidence of any systematic offsets in 
these data (\eg \citeANP{Harald98} 1998).

\begin{figure}
\centering
\centerline{\psfig{file=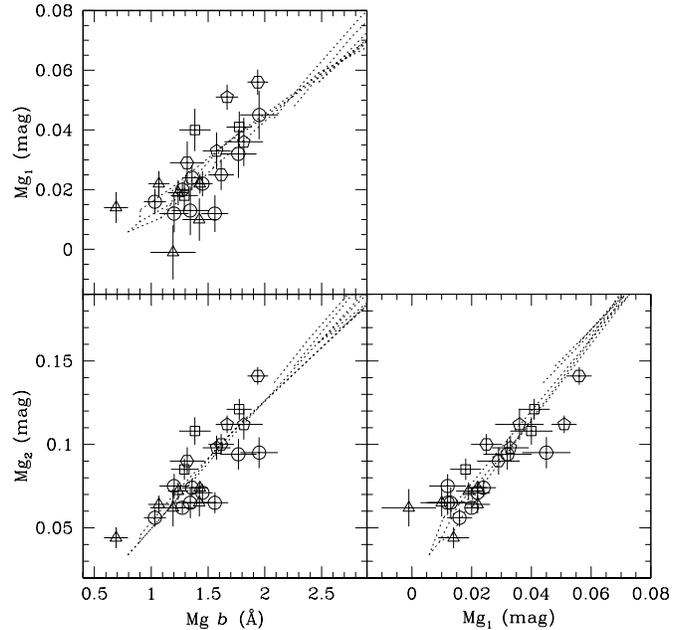,height=9cm}}
\caption{Lick/IDS magnesium indices of the LMC clusters, 
compared to the W94 stellar population models.
SWB-IVA clusters are indicated by triangles, IVB - squares, 
V - pentagons, VI - hexagons and SWB-type VII clusters 
are circles. These particular combinations of Lick/IDS indices
lead to the W94 model isochrones (dotted lines) to be 
superimposed onto lines of constant metallicity (\ie the models
are degenerate in age and metallicity). }
\label{fig:mg.indices}
\end{figure}

In Figure~\ref{fig:mg.indices} we show index-index
plots of the Mg {\it b}, Mg$_1$ and Mg$_2$ indices of the 
clusters, compared to the W94 models. We find that, despite
the apparent small range of metallicity covered by the
clusters due to the 'squashing' of the model grids, 
the agreement between the models and these data is good.
The absence of any significant offsets from the W94 models indicates 
that our resolution corrections, and the corrections derived from
the Lick standard stars are accurate. 

\begin{figure}
\centering
\centerline{\psfig{file=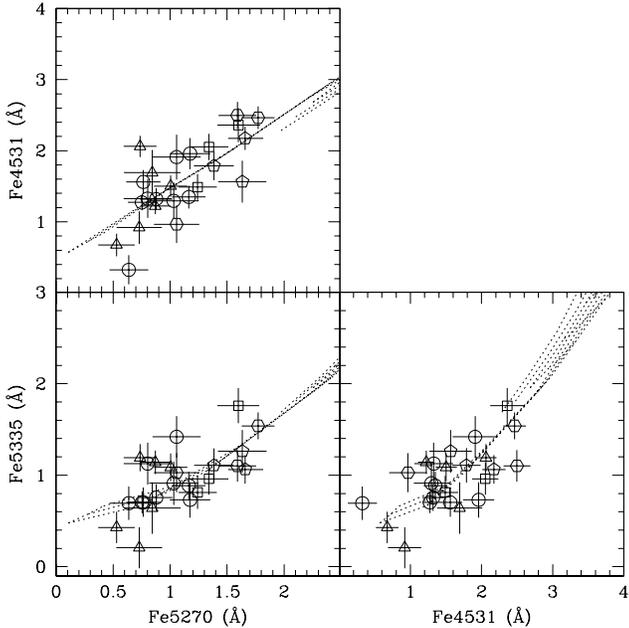,height=9cm}}
\caption{Lick/IDS iron indices of the LMC clusters. Symbols
as for previous figure.}
\label{fig:fe.indices}
\end{figure} 

The agreement between the iron indices
(Figure~\ref{fig:fe.indices}) is also generally good.
The scatter in the figure is somewhat larger than for
Figure~\ref{fig:mg.indices}, reflecting the greater 
statistical uncertainty in measuring these weaker features.

\begin{figure}
\centering
\centerline{\psfig{file=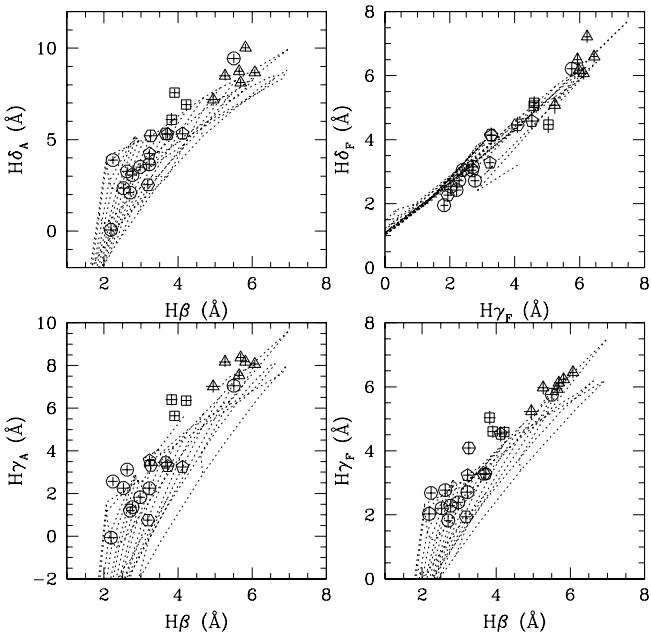,height=9cm}}
\caption{Lick/IDS Balmer indices of the LMC clusters, compared
to the models of Maraston \& Thomas (2000). Symbols as for 
Figure~\ref{fig:mg.indices}.}
\label{fig:balmer.indices}
\end{figure} 

In Figure~\ref{fig:balmer.indices} we compare the 
Balmer-line indices of the LMC clusters with the
models of \citeANP{Maraston00}. We do not use the
W94 models here since they do not reach to sufficiently
young ages.
In each of the panels in Figure~\ref{fig:balmer.indices}, ages
become progressively younger from the bottom-left to top right,
indicating that the LMC clusters possess a
significant range in age.
Notice that the panels in Figure~\ref{fig:balmer.indices}
which include H$\beta$ as one co-ordinate are not degenerate, 
demonstrating the increased age-sensitivity of H$\beta$
over the higher-order Balmer lines.
No significant offsets are apparent between these
indices.

\section{The Ages and Metallicities of the LMC Clusters}
\label{sec:TheAgesandMetallicitiesoftheLMCClusters}

\subsection{The Loci of SWB-Types on the SSP Grids} 
\label{subsec:TheLociofSWB-TypesontheSSPGrids} 

\begin{table}
\begin{center}
\caption[]
{The SWB -- age calibration of \citeANP{BCD92} 1992.}
\begin{tabular}{ll} 
\hline
SWB Type & Age (Gyr)\\
\hline 
I	&	0.01 --	0.03\\
II	&	0.03 --	0.07\\
III	&	0.07 -- 0.20\\
IVA	&	0.20 -- 0.40\\
IVB	&	0.40 -- 0.80\\
V 	&	0.80 -- 2.00\\
VI	&	2.00 -- 5.00\\
VII	&	5.00 -- 16.00\\
\hline
\end{tabular}
\label{tab:swb.ages}
\end{center}
\end{table}

In principle, the position of any star cluster in the 
age-metallicity plane of the SSP grids indicates the
age and metallicity of that stellar population.
As discussed in $\S$~\ref{subsec:StarClustersintheLMC}, 
SWB80 grouped the LMC clusters into SWB-types, depending
upon their {\it Q(ugr)-Q(vgr)} colours. Since the SWB ranking
is effectively one of age, this should be reflected in the locus
of the SWB types on the SSP grids. For the convenience of the  
reader, we reproduce the age groups for 
SWB types I--VII in Table~\ref{tab:swb.ages}, as given by the
calibration of \citeANP{BCD92} (1992).

\begin{figure*}
\centering
\centerline{\psfig{file=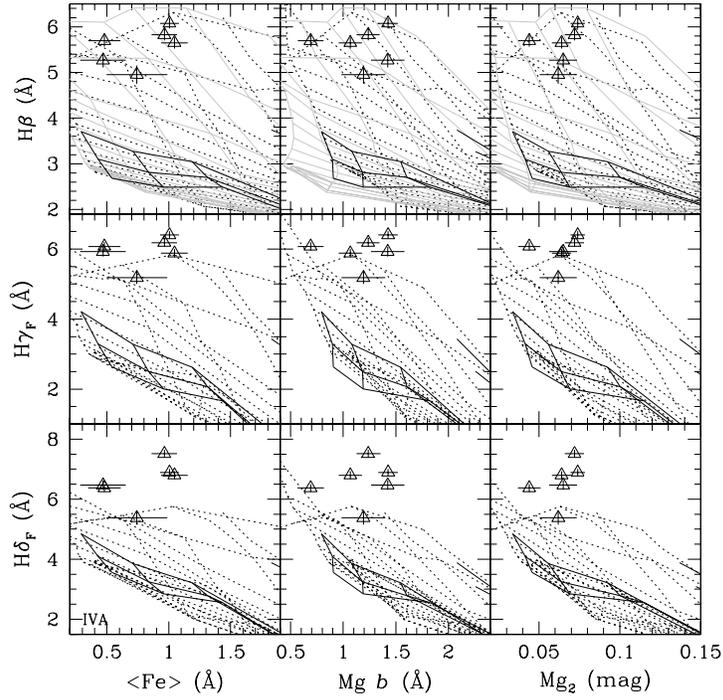,height=10cm}}
\caption{Age-metallicity diagnostic diagrams for the
SWB IVA clusters. {\it Grey lines:} Lick/IDS indices
of the LMC clusters compared to the KFF99 SSP 
models. Near-horizontal lines are isochrones 
of 0.5, 1, 2, 3 ... 16 Gyr (youngest ages at the top).
Metallicity increases from left to right, with lines
of --2.3, --1.7, --1.22 (interpolated), --0.7, --0.4, 0 
and + 0.4 dex. 
{\it Dotted lines:} SSP models of Maraston \& Thomas (2000).
Isochrones range from 0.5, 1, 2, 3 ... 15 Gyr, 
metallicity isopleths are --2.25, --1.35, --0.84 (interpolated),
--0.33, 0 and + 0.35 dex.
{\it Solid lines:} Worthey (1994) models with ages 8, 12 
and 17 Gyr, and metallicities (shown) --2.0, -1.5 and --1.0.}
\label{fig:swbIVA}
\end{figure*}

\begin{figure*}
\centering
\centerline{\psfig{file=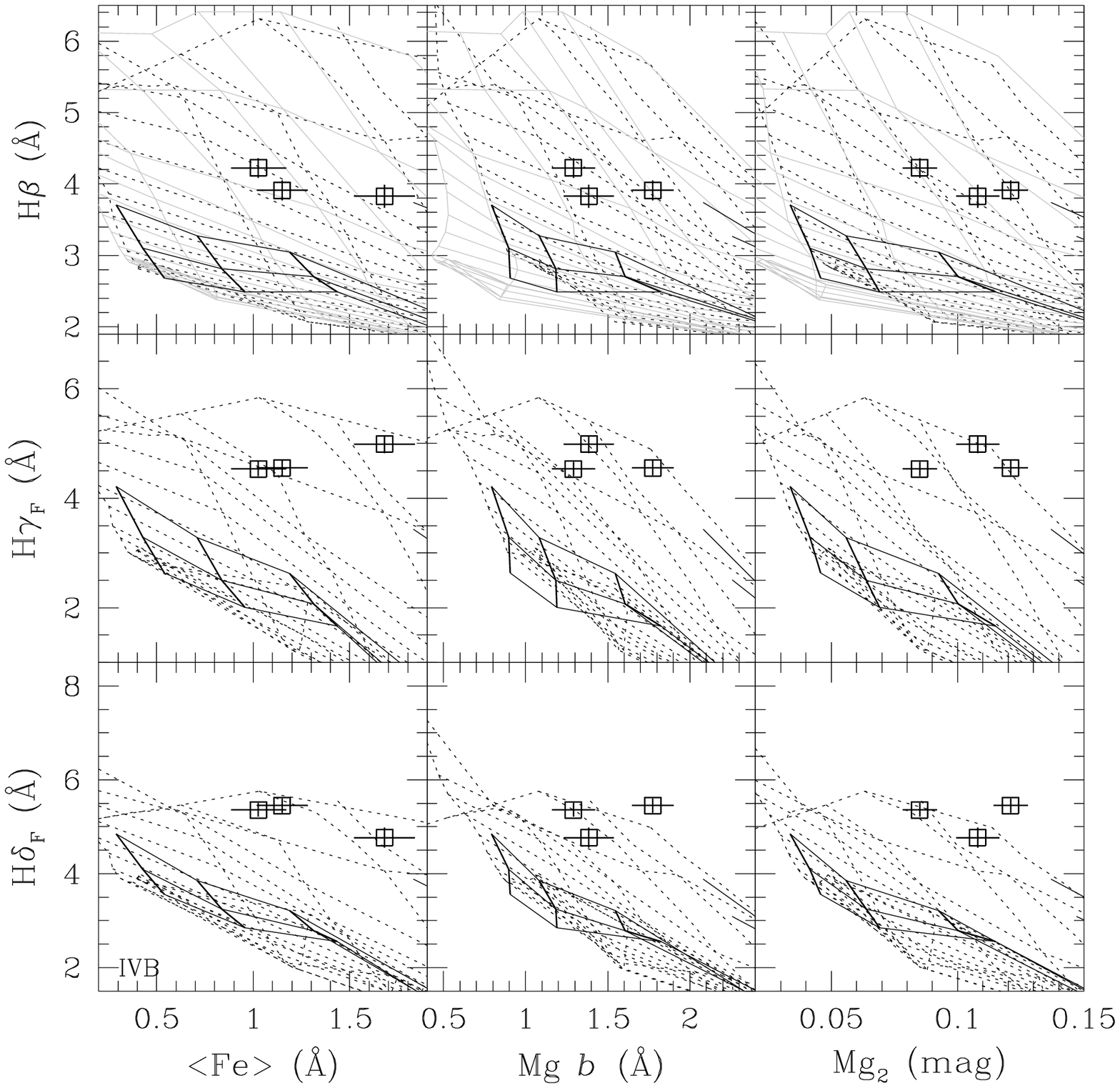,height=10cm}}
\caption{Age-metallicity diagnostic diagrams for the
SWB IVB clusters.}
\label{fig:swbIVB}
\end{figure*}

\begin{figure*}
\centering
\centerline{\psfig{file=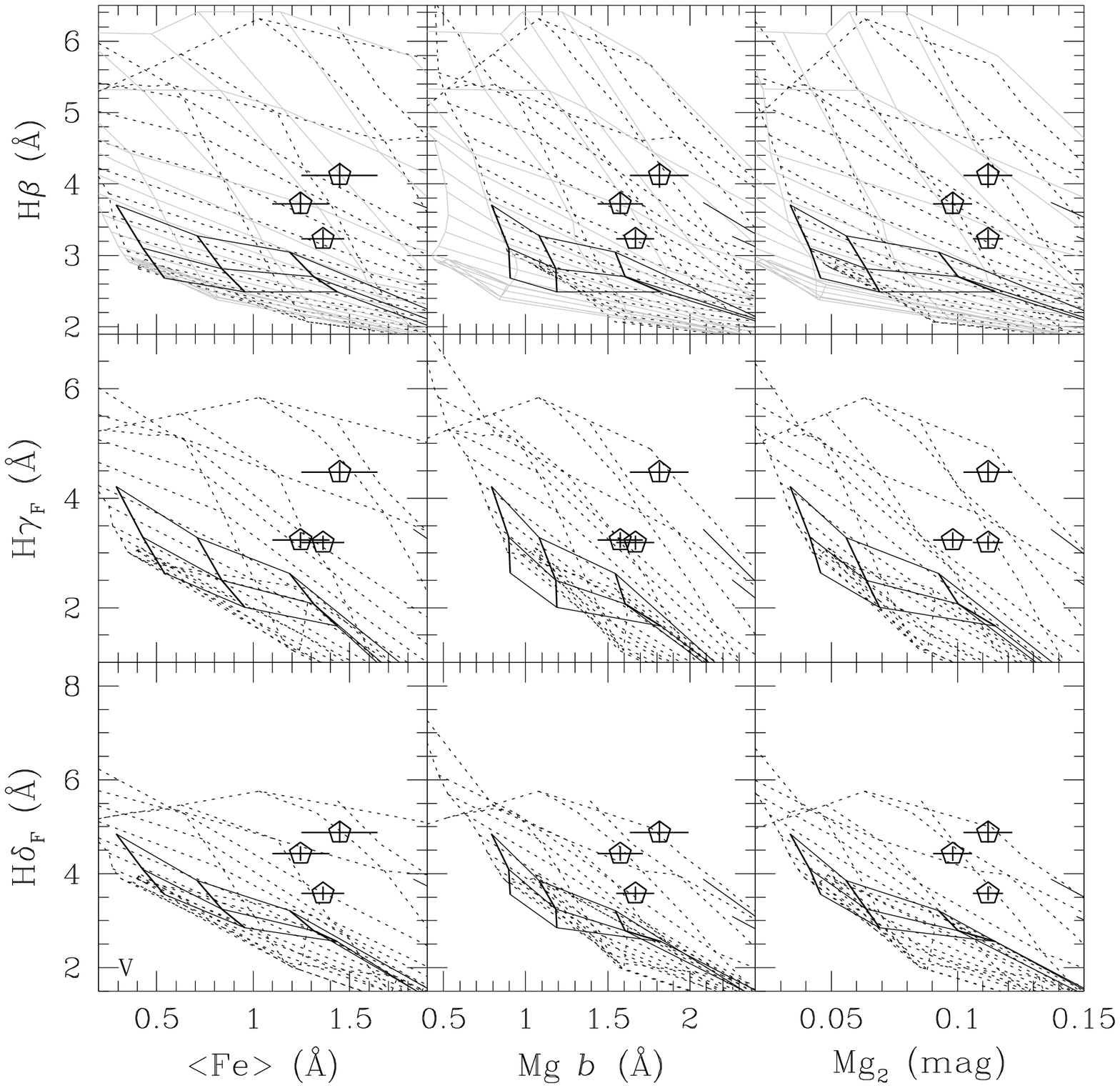,height=10cm}}
\caption{Age-metallicity diagnostic diagrams for the
SWB V clusters.}
\label{fig:swbV}
\end{figure*}

\begin{figure*}
\centering
\centerline{\psfig{file=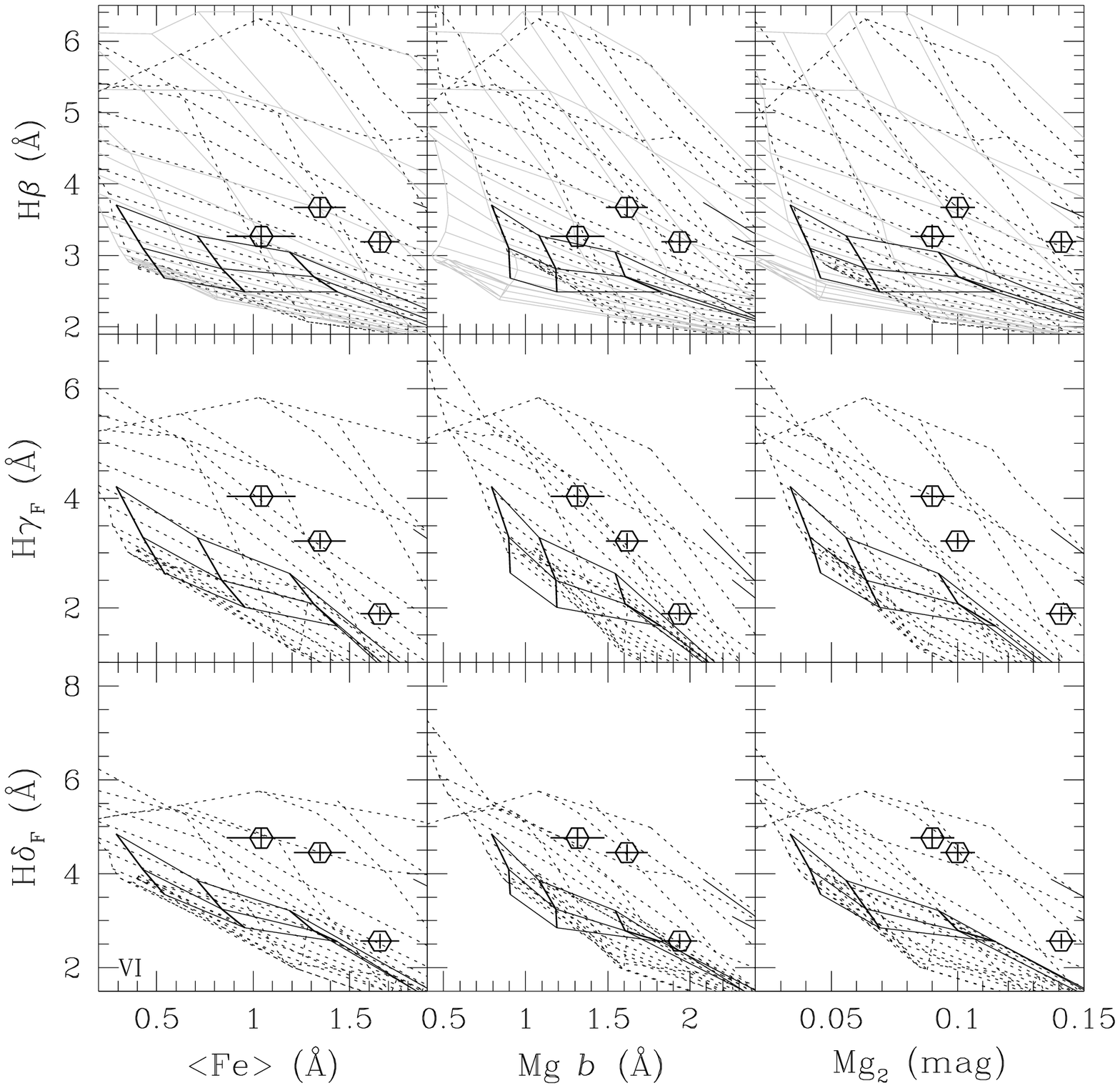,height=10cm}}
\caption{Age-metallicity diagnostic diagrams for the
SWB VI clusters.}
\label{fig:swbVI}
\end{figure*}

\begin{figure*}
\centering
\centerline{\psfig{file=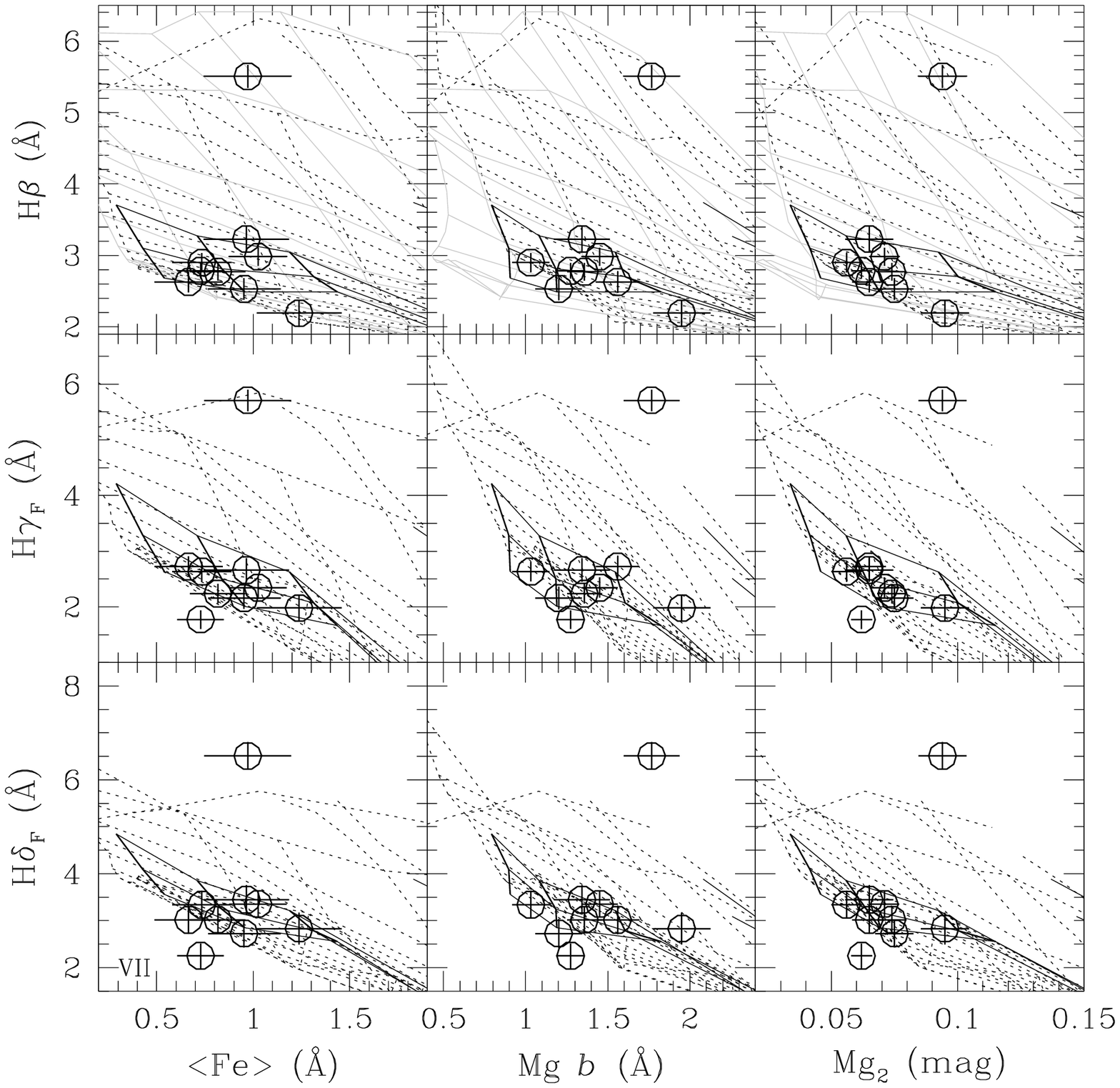,height=10cm}}
\caption{Age-metallicity diagnostic diagrams for the
SWB VII clusters.}
\label{fig:swbVII}
\end{figure*}

In Figures~\ref{fig:swbIVA} to~\ref{fig:swbVII}, we show the
metallicity-sensitive $\langle$Fe$\rangle$, Mg $b$ and Mg$_2$ indices
versus the more age-sensitive H$\beta$, H$\gamma_F$, and
H$\delta_F$ indices of the LMC clusters. In each case, 
we compare them to the SSP models of KFF99, 
\citeANP{Maraston00} and W94. 

Our goal is to empirically test the age and metallicity
predictions of SSP models which predict line-strength
indices in the Lick/IDS system, rather than
provide an exhaustive comparison between
different models (\eg see \citeANP{Maraston01a} 2001).
However, there are clearly significant differences between the 
SSP models, which warrant some discussion before using them
to derive ages and metallicities for the clusters.

Perhaps the most noticeable  difference between the models 
is the much smaller parameter space covered
by the W94 models As discussed in
$\S$~\ref{subsec:TheSSPModels}, the majority
of the LMC clusters do not fall
onto the W94 grids, indicating that the W94 models
will only be useful for the oldest clusters.
Direct comparison between the KFF99 and \citeANP{Maraston00}
models shows that, in general, the KFF99 models
tend to predict younger ages and higher metallicities
for any given position on the grids. Also, 
the Mg {\it b} and Mg$_2$ indices predicted by the
KFF99 models extend to lower values than those
of \citeANP{Maraston00}, despite the fact that both models
extend to [Fe/H] $\sim$ --2.3.

One feature of the \citeANP{Maraston00}  models 
is that, at low metallicities, the youngest isochrones 
turn sharply downwards, crossing over older isochrones.
This is a result of Balmer line-strengths decreasing as turn-off
temperatures exceed T$_{\rm eff} \sim$ 9,500 K, at younger ages
than where Balmer lines are at a maximum.
The loci of the Balmer-line maximum shifts to older ages
as metallicity decreases (\eg \citeANP{Maraston01} 2001).

An important, but subtle difference
between the evolutionary synthesis models of 
\citeANP{Maraston00} and KFF99, and the population
synthesis models of W94, is the treatment of RGB stars.
In the \citeANP{Maraston00} and KFF99 models, 
mass-loss on the RGB is dealt with by
using an analytical recipe (in the case of these
two SSP models, Reimers' mass-loss equation \cite{Reimers75}
is adopted).
The amount of mass-loss suffered by RGB stars
plays a crucial r$\hat{\rm o}$le in determining
the position at which these stars fall onto the
HB (\eg \citeANP{Buzzoni89} 1989). Lower mass
RGB stars lead to bluer (hotter) HBs, and 
increasing the 'mass-loss parameter'
($\eta$), pronounces this effect.
As pointed out by \citeANP{Rabin82} (1982), the presence
of such stars can potentially severely effect integrated indices, 
and in particular the Balmer lines.

The effect of this mass-loss can be 
seen in the \citeANP{Maraston00} models, and to a lesser 
extent, those of KFF99. At old ages ($\sim$ 12 Gyr), and 
low metallicities ([Fe/H] $\leq$ --1.0), 
Balmer line-strengths begin to $increase$, 
rather than  decrease as expect with age, 
creating a saddle-point minima at $\sim$ 12 Gyr.
Therefore, at old ages and low 
metallicities, a unique solution for 
the age of a stellar population
is not necessarily attainable with integrated
indices. The interpretation of the
SSP models in this regime is further discussed
in $\S$~\ref{subsec:AgeComparisons}, as are the
effects of HB stars on our integrated
Balmer indices in $\S$~\ref{subsubsec:HorizontalBranchStars}.

Returning to Figures~\ref{fig:swbIVA} 
to~\ref{fig:swbVII}, we find the agreement between 
the $\langle$Fe$\rangle$, Mg $b$ and Mg$_2$
indices is good. 
In general, the position of the clusters on the SSP
grids using these different metallicity indicators are
consistent within the uncertainties.
The H$\beta$, H$\delta_{\rm F}$ and H$\gamma_{\rm F}$ indices 
also behave similarly, although there are some indications
of systematic differences between their age predictions 
for the younger clusters.

The SWB types lie in relatively
tight groups on the SSP model grids.
As one goes to older SWB types in Figures~\ref{fig:swbIVA} 
to~\ref{fig:swbVII}, these groups trace a characteristic
shape, moving from the top-left (young ages, 
[Fe/H] $\gsim$ --1.0), to centre-right (intermediate
ages, [Fe/H] $\gsim$ --1.0), to bottom-left (old ages, 
[Fe/H] $<$ --1.0).
One cluster in Figure~\ref{fig:swbVII} clearly stands 
out as being significantly younger than the other
SWB VII clusters, and provides a nice illustration
of the advantage of integrated spectroscopy
over photometry in separating age and metallicity
effects.
This cluster, NGC~1865, has an SSP-derived
age of $<$ 1.0 Gyr, rather than the $\sim$ 10 Gyr
implied by its SWB type.
\citeANP{Geisler97} (1997) obtained an age 
of 0.9 Gyr for this cluster from the position
of its CMD turnoff, which is consistent with our 
value. \citeANP{Geisler97} (1997) attributed
its SWB mis-classification to a combination of
the stochastic effects of bright stars in the cluster, 
and its relative faintness contrasted with a dense 
stellar background.

\subsection{Measuring Metallicities and Ages}
\label{subsec:MeasuringMetallicitiesandAges}

As our principle metallicity indicators we
have chosen our best-measured 
magnesium dominant Lick index 
(Mg$_2$), and mean of two iron indices
(Fe5270 and Fe5335) 
in the form of $\langle$Fe$\rangle$.
As age-sensitive indices, we use H$\beta$ 
(\citeANP{Maraston00} and KFF99 models),
in addition to our best-measured higher-order Balmer lines
H$\gamma_{\rm F}$ and H$\delta_{\rm F}$
(\citeANP{Maraston00} models only).

Ages and metallicities are obtained 
for each cluster by interpolating the
model grids using a \textsc{fortran} 
programme. 
Where clusters lie off the grids, linear extrapolation is used.
Uncertainties are derived by
perturbing the line-strength indices
by their corresponding measurement error.
Because of the non-orthogonal nature of the
model grids, two differing uncertainties in age
and two differing uncertainties in metallicity are 
obtained.  
As discussed previously, due to the effects of the HB
on the Balmer indices, the metal-poor, old regions models often have
two age solutions for the SWB VII clusters.
In these cases, we adopt the prior that these clusters
are Galactic GC analogues, and adopt the older 
ages if the clusters are consistent with these old isochrones 
(see $\S$~\ref{subsec:AgeComparisons} for futher explanation.)

Whilst we wish to compare the SSP model predictions 
to {\it independently} derived ages and metallicities, 
an important feature of the SSP models should be emphasised.
Due to the non-orthogonality of the SSP 
grids, changes in the metallicity estimates of the 
clusters effect the derived  ages of the 
clusters and {\it vice versa} -- \ie\ there is still
an age-metallicity degeneracy.
As a direct result of this, the KFF99 models, which have a 
tendency to systematically over-predict metallicities
with respect to the \citeANP{Maraston00} models 
(see $\S$~\ref{subsec:TheLociofSWB-TypesontheSSPGrids}),
systematically under-predict the cluster ages with respect
to the \citeANP{Maraston00} models.

\begin{figure*}
\centering
\centerline{\psfig{file=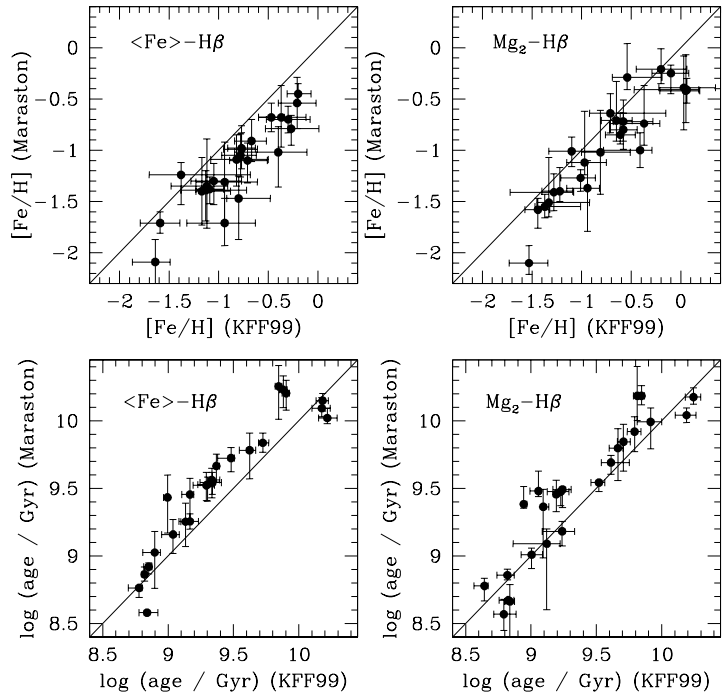,height=10cm}}
\caption{Comparison between the age and metallicity predictions
of the Maraston \& Thomas and KFF99 SSP models for the
$\langle$Fe$\rangle$, Mg$_2$
and H$\beta$ Lick/IDS indices. Solid line represents unit slope.}
\label{fig:compare.ssp}
\end{figure*}

This is illustrated in 
Figure~\ref{fig:compare.ssp}, where we compare the 
age and metallicity predictions of the \citeANP{Maraston00} 
and KFF99 models using two different metallicity
indicators ($\langle$Fe$\rangle$, Mg$_2$) and
the more age-sensitive H$\beta$. 
The effect can be most clearly seen
in the $\langle$Fe$\rangle$--H$\beta$ plane
of the models. The KFF99-derived metallicities
are systematically 0.2--0.5 dex higher than the 
\citeANP{Maraston00}-derived metallicities. This leads 
to cluster ages which are significantly younger
in the KFF99 models (up to 6 Gyr for old ages) than
those of \citeANP{Maraston00}.
However, surprisingly, the agreement between models
for the ages of the youngest clusters is much better, even though
the agreement between their metallicities is poorest. The origin
of these differences are unclear, since both the models
use the same fitting-functions, but a possible 
explanation may lie in their adoption of different
input isochrones.

Of final, important note in Figure~\ref{fig:compare.ssp}, 
there is a clear offset in metallicity between the $\langle$Fe$\rangle$-
and Mg$_2$-derived metallicities, in the sense that the
Mg$_2$ index predicts metallicities 0.1 $\sim$ 0.5
dex higher than $\langle$Fe$\rangle$. 
However, we find no evidence of a significant 
systematic offset between our measured magnesium and iron
indices. As we showed in Section~\ref{sec:TheSpectroscopicSystem},
we were able to correct both these indices onto the 
Lick/IDS system. 

An alternative explanation is that the $\langle$Fe$\rangle$ and 
Mg$_2$ indices do not track metallicity in the same manner; the 
[Fe/H] measurements of the clusters are 
systematically lower than our [Mg/H] measurements. 
Such "$\alpha$-enhancement" has been seen in
the integrated spectra of elliptical galaxies 
(e.g. \citeANP{Peletier89} 1989; 
Gonz$\acute{a}$lez~1993; \citeANP{Harald01} 2001;
\citeANP{Trager00} 2000a) and recently in extragalactic
globular clusters (\eg \citeANP{Forbes01b} 2001, 
\citeANP{Larsen02} 2002). 
Moreover, high-resolution spectroscopy of LMC
clusters giants suggests that, from [O/H] ratios, this 
is also the case for LMC clusters \cite{Hill00}.
However, such an interpretation is complicated by the fact that,
at low metallicities, the differences between solar-scaled and
$\alpha$-enhanced SSP models are small (\ie little dynamic range,
\eg \citeANP{Milone00} 2000).
A detailed analysis of this important issue is beyond
the scope of this paper, and we defer further discussion to
future work.

In Table~B1, we list the age and metallicity 
predictions of the \citeANP{Maraston00} SSP models, 
using the Mg$_2$-H$\beta$ and 
$\langle$Fe$\rangle$-H$\gamma_{\rm F}$ indices\footnote{For a 
full list of each of the index-index-model 
predictions see {\bf http://astronomy.swin.edu.au/staff/mbeasley/lmc/}.}.
To test these model predictions, we have collected age
and metallicity estimates for the LMC clusters in our sample
from the literature. 
These are also presented in Table~B1.
These literature ages and metallicities come from
a variety sources and are therefore rather inhomogenous.
Where possible, we have tried to minimise this inhomogeneity 
whilst retaining a large enough sample for meaningful comparisons.
Cluster ages are preferentially taken from studies which
have located the main sequence turn-off in CMDs. These have been
supplemented with ages obtained by the giant-branch calibration 
of \citeANP{Mould82} (1982). For the seven clusters with no
previous spectroscopic- and/or CMD-derived age determinations, 
we have adopted a mean age
which corresponds to their SWB type (see
Table~\ref{tab:swb.ages}). We assign uncertainties by taking the
50 percentiles in their SWB age-range.

The majority of the literature cluster metallicities come from 
the $\sim$ 2 \AA\ resolution Ca-triplet spectroscopy of cluster
red giants by \citeANP{Olszewski91} (1991). 
For clusters not in the \citeANP{Olszewski91} (1991) sample, 
we adopt metallicities from integrated spectroscopy
(\citeANP{Rabin82} 1982; \citeANP{Dutra99} 1999).
Ages and metallicities for NGC~1806 are from 
the Str\"{o}mgren photometry of \citeANP{Dirsch00} (2000).

The LMC globular clusters NGC~1754, NGC~1835, NGC~1898, NGC~2005
and NGC~2019 have metallicities 
from two sources. 
\citeANP{Olsen98} (1998) derived metallicities
for these clusters from \HST\ CMDs by measuring
the height of the HB above the main sequence
turn-off (the $V^{TO}_{HB}$ method of \citeANP{Sarajedini94}
1994). These clusters are also included in the
\citeANP{Olszewski91} (1991) sample, and therefore
have spectroscopic abundances.
In the mean, the metallicities 
derived by \citeANP{Olszewski91} (1991) are $\sim$ 0.3 dex
lower than those found by \citeANP{Olsen98} (1998) using
the $V^{TO}_{HB}$ method.
To increase the homogeneity in our metallicity comparisons, 
and because we prefer spectroscopic metallicities, we
adopt the values of \citeANP{Olszewski91} (1991), and the
ages derived by \citeANP{Olsen98} (1998) using these
metallicities.
NGC~1916 was also in the \citeANP{Olsen98} (1998) sample, 
however differential reddening precluded an age determination
for this cluster.

\subsection{Metallicity Comparisons}
\label{subsec:MetallicityComparisons}

\begin{figure*}
\centering
\centerline{\psfig{file=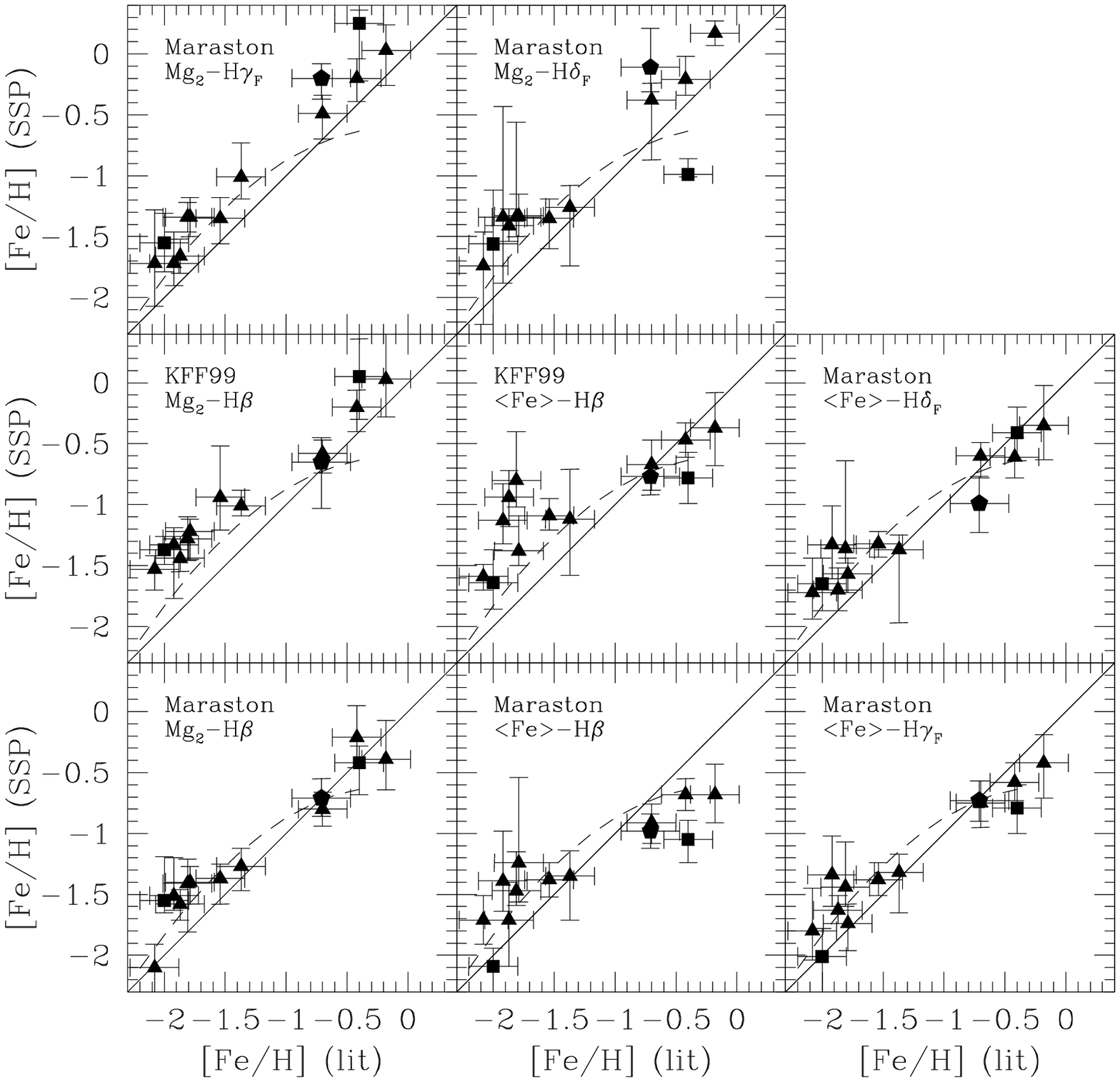,height=18cm,width=18cm}}
\caption{Comparison between literature metallicities of the LMC
clusters, and those derived from the SSP models of Maraston \&
Thomas and KFF99 for different spectroscopic indices. 
Solid triangles represent the spectroscopic
values of Olszewski \etal (1991), squares : integrated
spectroscopy, pentagons : Str\"{o}mgren photometry.
The solid line indicates unit slope, dashed line
indicates the non-linear relation between the Zinn \&
West (1984) and Zinn (1985) metallicity scale, and that obtained by
Carretta \& Gratton (1997).}
\label{fig:feh}
\end{figure*}

We compare the metallicities of the LMC clusters derived from
the various combinations of metallicity-sensitive and
age-sensitive indices (using the \citeANP{Maraston00} 
and KFF99 models)
with their literature values in Figure~\ref{fig:feh}. 
To first order, the agreement between the literature values
and the metallicities from the $\langle$Fe$\rangle$ and
Mg$_2$ indicators is satisfactory. 
Our mean metallicity uncertainty is
$\sim$ 0.20 dex for Mg$_2$--H$\beta$ and $\sim$ 0.25 dex
for $\langle$Fe$\rangle$--H$\beta$, generally 
consistent with the scatter seen in Figure~\ref{fig:feh}.
For the higher-order Balmer lines, the mean uncertainty
increases to $\sim$ 0.3 dex due to the larger
degree of age-metallicity degeneracy in the models
for these indices.

There is broad agreement between different index-index
combinations, as there is reasonable
consistency between the \citeANP{Maraston00} and KFF99 
model metallicity predictions (for H$\beta$).
The well documented metallicity-gap (and the corresponding
age gap) of the LMC clusters is evident at [Fe/H] $\sim$ --1.0
(\eg \citeANP{Westerlund97} 1997).

\begin{figure*}
\centering
\centerline{\psfig{file=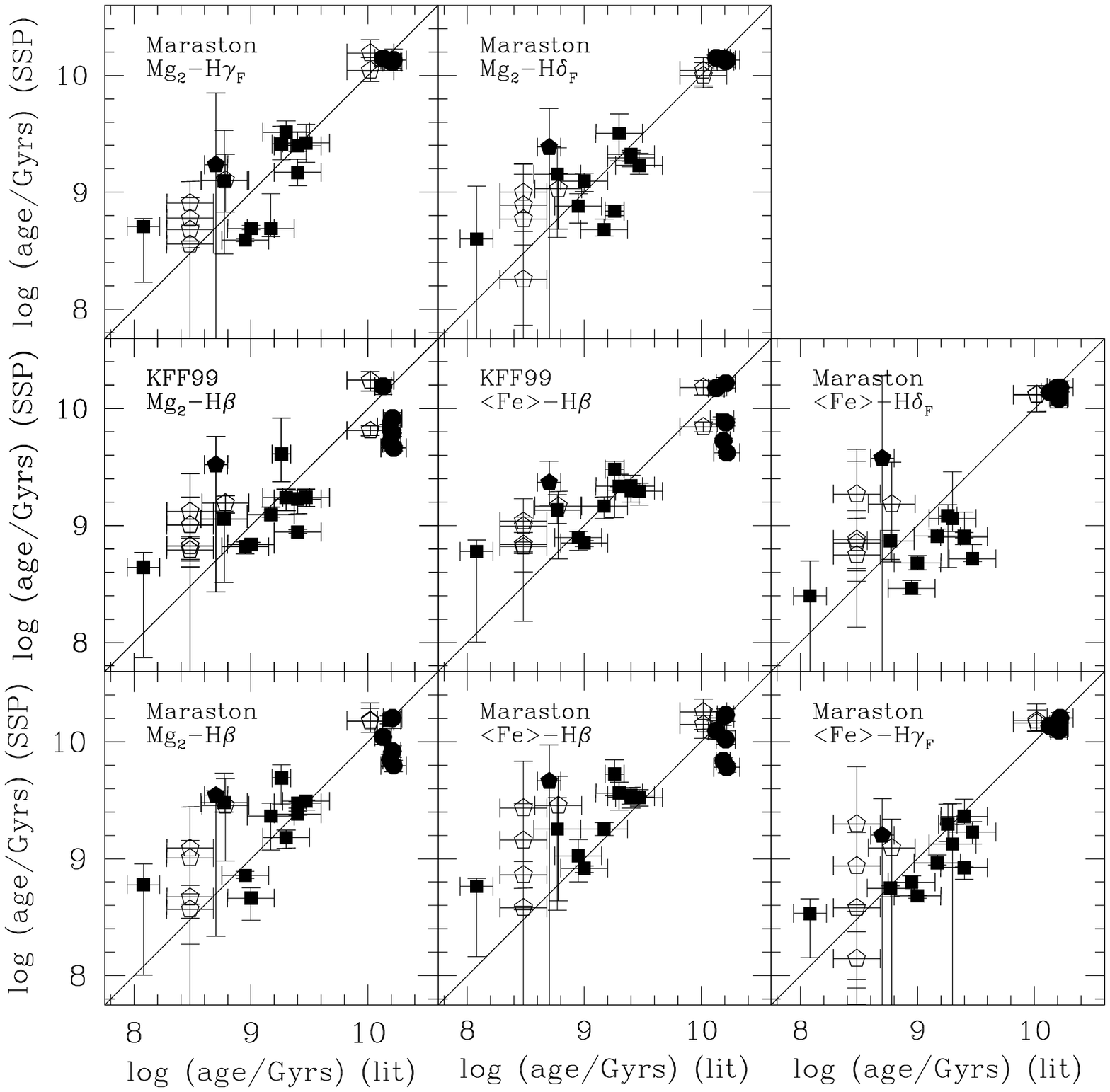,height=18cm,width=18cm}}
\caption{Comparison between literature ages of the LMC
clusters, and those derived from the SSP models of 
Maraston \& Thomas and KFF99. Solid squares indicate literature ages from
ground-based CMDs, pentagons : Str\"{o}mgren photometry, circles
: \HST\ CMDs, open pentagons : ages from SWB types. The solid
line represents unit slope.}
\label{fig:age}
\end{figure*}

However, closer inspection of Figure~\ref{fig:feh} reveals
some interesting differences between the SSP-derived and 
literature metallicities. The most obvious difference is that 
again, the Mg$_2$ index predicts higher metallicities than 
$\langle$Fe$\rangle$. As discussed in the previous Section, in
the absence of any obvious systematic errors in our data, a
possible explanation is the detection of $\alpha$-enhancement in
the LMC clusters.

Another discrepancy is that the low-metallicity clusters 
generally exhibit a systematic deviation from unit slope. 
Rather than showing a straight offset from the literature values, 
these clusters metallicities are actually better fit
with a non-linear function. 
This behaviour is also seen to greater 
or lesser extents in all the other indices
and in the both the KFF99 and \citeANP{Maraston00} models.

Can the origin of these differences lie in the different metallicity
scales of the literature LMC clusters, and the SSP models? 
The source of the majority of our literature cluster 
metallicities come from the study of \citeANP{Olszewski91} (1991).
\citeANP{Olszewski91} (1991) calibrated 
their Ca-triplet measurements with Galactic GCs using  
metallicities on the \citeANP{Zinn84} (1984) and 
\citeANP{Zinn85} (1985; hereafter
the ZW scale) scale. However, \citeANP{Carretta97} (1997)
have shown that, whilst the ZW scale is reassuringly
monotonic, it is non-linear with respect to the solar (meteoritic) scale
upon which the SSP model input isochrones are based.
In the range --1.9 $\leq$ [Fe/H] $\leq$ --1.0, the
ZW scale predicts metallicities $\sim$ 0.2 dex lower
than the meteoritic values. For metallicities 
outside this range, the ZW scale yields values systematically
higher than the meteoritic scale.

The relation between these two scales is illustrated
in Figure~\ref{fig:feh} by the dashed line. The $x$-axis
may be regarded as the ZW scale (largely \citeANP{Olszewski91}
metallicities), the $y$-axis that 
of \citeANP{Carretta97} (1997) (the SSP models).
We find that, for the $\langle$Fe$\rangle$
metallicity indicator, better agreement 
is generally reached using the \citeANP{Carretta97} (1997) scale, 
including the flattening in their relation 
at [Fe/H] $\sim$ --0.5. This is perhaps not surprising since 
\citeANP{Carretta97} (1997) derived metallicities from resolved 
Fe I lines, whilst  $\langle$Fe$\rangle$ essentially
measures features (iron and others) in the same wavelength range 
but at significantly lower resolution.
However, the metallicities derived using Mg$_2$ are not
so well reconciled, particularly with regard to the
highest-metallicity clusters which actually follow
a one to one correlation (unit slope) somewhat more closely. 
The question of which metallicity scale is to be 
preferred is still an open one (\eg see \citeANP{Caputo02} 2002).

To summarise, we find good agreement between the 
SSP model and literature determinations for the metallicities 
of the LMC clusters in the range --2.0 $\leq$ [Fe/H] $\leq$ 0.
Systematic offsets between the metallicities predicted by
$\langle$Fe$\rangle$ and Mg$_2$ are possibly due to 
non-solar abundance ratios in the LMC clusters.
Differences in the metallicity scales (from the line-blanketing
indices of ZW, and Fe I scale of \citeANP{Carretta97} 1997)
are at most a second-order effect.

Of the 24 LMC clusters in our sample, 11
have no previous metallicity determinations. 
The weighted mean of the two values given 
in Table~B1 should yield a good mean metallicity
for these clusters.

\subsection{Age Comparisons}
\label{subsec:AgeComparisons}

Comparisons between the ages of the clusters derived from
the SSP models, and those collected from the literature
are presented in Figure~\ref{fig:age}.
The presence of clear correlations in the 
figure are reassuring. Each of the Balmer indices
shown in Figure~\ref{fig:age} are, to first order
at least, tracking the temperature
of the main sequence turn-off in the clusters, 
and thereby providing a measure of 
cluster age.

We find that the best agreement is obtained 
for the intermediate-aged clusters (\ie 1 $\sim$ 4 Gyr). 
The H$\beta$ index yields ages which are 
in excellent agreement with their CMD values, 
for both the iron and magnesium metallicity 
indicators. H$\gamma_{\rm F}$ and 
H$\delta_{\rm F}$ also predict ages which 
are generally consistent with the literature, 
although with somewhat larger scatter.

In the metallicity--H$\beta$ planes of both
models, the youngest clusters ($<$ 1 Gyr) are 
predicted to be older (by 0.2 -- 1.0 Gyr) than the literature
ages. These H$\beta$ ages are also older than is 
predicted by the higher-order Balmer lines. 
Whilst this is only a 1--2$\sigma$ effect
for any individual cluster, it is systematic.
In the absence of systematic errors in the 
literature ages of these clusters (in the sense that the
literature ages are too young), the source of this 
disagreement is either
{\it (i)} emission fill-in of the clusters'
H$\beta$ lines or {\it (ii)} an uncertainty in the
SSP models at young ages.
Since we detect no emission in these clusters'
sky spectra, we conclude
that, if present, any emission must be arising from some 
internal source. This would be plausible if these
clusters were very young (\ie $<$ 10$^{7}$ yr) and massive
O/B stars were present. 
However, this is not only inconsistent with their
integrated colours, but also the spectra of these clusters do not show 
the characteristic blue continua of very young objects. 

There is some evidence that 
the SSP models at these young ages 
may be at fault (for H$\beta$).
The disagreement in the cluster ages occurs at $<$ 3 Gyr, coincident
with were the SSP models are most uncertain due
to the necessary extrapolation required in the Lick/IDS 
fitting-functions.
Moreover, the young, metal-poor regions
of the SSP grids are precisely the areas of parameter
space which are inadequately covered by the 
spectral libraries input into the models.
Clearly, this issue 
needs to be investigated futher with a larger
sample of integrated spectra for Magellanic Cloud 
clusters in this age range.

We now turn to the SSP age predictions of the
LMC globular clusters (SWB VII) in our sample.
At this stage, the method by which we 
interpret the ages of the 'old' clusters 
using the SSP model grids should be discussed. 

\begin{figure}
\centering
\centerline{\psfig{file=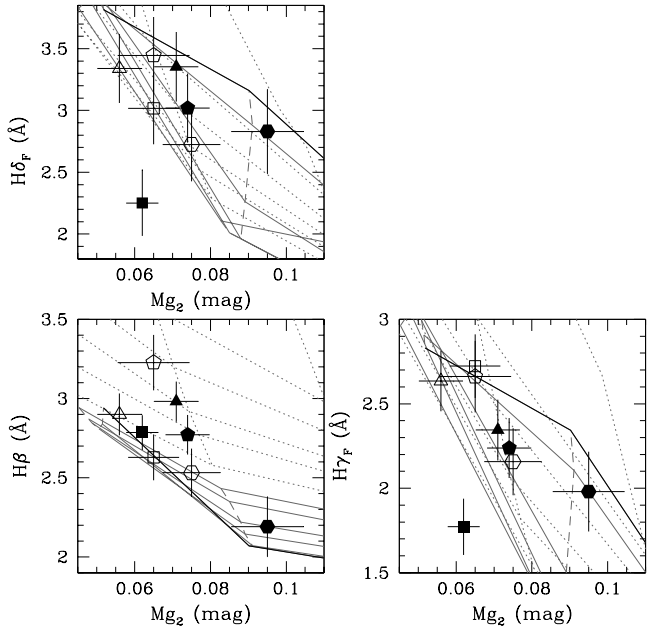,height=9cm}}
\caption{The LICK/IDS indices of the LMC globular 
clusters compared to the SSP models of Maraston \& Thomas (2000).
Symbols indicate the different clusters, 
filled triangle : NGC~1754, filled square : NGC~1786, 
filled pentagon : NGC~1835, filled hexagon : NGC~1898,
open triangle : NGC~1916, open square : NGC~1939, 
open pentagon : NGC~2005, open hexagon : NGC~2019. 
Dotted grey lines indicate isochrones of 5--9 Gyr, 
solid grey lines are isochrones of 10--14 Gyr.
The solid black line indicates the position of the
oldest (15 Gyr) isochrone of the models.
Dashed line indicates an interpolated line of constant metallicity
at [Fe/H]=--1.22.}
\label{fig:old.globulars}
\end{figure}

In Figure~\ref{fig:old.globulars} we show
the SWB VII clusters (excluding NGC~1865 which
we earlier showed to be $\sim$ 1.0 Gyr old) in the Mg$_2$--Balmer
line planes of the \citeANP{Maraston00} models.
It can be seen immediately from Figure~\ref{fig:old.globulars}
that the isochrones of the oldest ages overlap
with those of younger isochrones at low metallicities.
Whilst this is only a weak effect in the models for
H$\beta$, for H$\gamma_{\rm F}$ and 
H$\delta_{\rm F}$ the effect is significant.
The oldest (15 Gyr) isochrone of the \citeANP{Maraston00} 
models predicts higher-order Balmer line-strengths
very similar to those of the 5 Gyr isochrone.

Our prior is such that the SWB VII clusters
shown in Figure~\ref{fig:old.globulars}
are the LMC counterparts of Galactic GCs, 
and as such have ages in excess of 10 Gyr.
Therefore, in interpreting the figure, 
the older isochrones ($>$ 10 Gyr) are
adopted. These are the cluster ages listed
in Table~B1, and shown in Figure~\ref{fig:age}.
Any clusters which lie {\it above} the oldest
(15 Gyr) isochrone of the models are assigned 
an extrapolated age (\ie $>$ 15 Gyr) only {\it if} their
line-strengths are consistent (1$\sigma$ uncertainty)
with this oldest isochrone (solid line
in Figure~\ref{fig:old.globulars}).
Otherwise, SWB VII clusters with strong
Balmer-absorption are given younger 'spectroscopic'  ages.

The age predictions of the SSP models for the 
LMC globular clusters (Figure~\ref{fig:age})
show significant variations, depending upon the 
combination of Lick/IDS indices used. 
In contrast, the literature ages of the 
clusters (5 from the \HST\ CMDs of \citeANP{Olsen98} (1998), 
1 from the ground-based CMD of \citeANP{Geisler97} (1997)
and 2 inferred from their SWB type) show
a reasonably tight age-range, from 10 to 17 Gyr.
These ages are in reasonable agreement with the 'best'
value for the Milky Way GCs of 12.9 $\pm$ 2.9 Gyr
\cite{Carretta00}.

We find that the higher-order Balmer lines, H$\gamma_{\rm F}$ and 
H$\delta_{\rm F}$, are generally consistent with 
the CMD turn-off ages and/or integrated colours.
However, H$\beta$, the most age-sensitive of the 
Lick/IDS indices \cite{WO97} predicts a couple of clusters to 
be significantly {\it younger} than is indicated by their
literature values.
Specifically, the two clusters, NGC~1754 and NGC~2005, 
have SSP model ages which are inconsistent with the 
literature at 3$\sigma$ significance. Both of these clusters
have well-developed blue HBs, and their
influence upon the integrated Balmer indices
we examine shortly.

As shown in  Figure~\ref{fig:old.globulars}, we
also find that the position of NGC~1786 
stands out in the H$\gamma_{\rm F}$ and H$\delta_{\rm F}$ 
planes of the SSP models.
However, these indices of this cluster 
only fall below the \citeANP{Maraston00} model isochrones
at the $\sim$ 2$\sigma$ level, and therefore we do not  
ascribe them any particular significance.

\subsubsection{The Effect of Horizontal Branch Stars on 
Integrated Balmer Indices}
\label{subsubsec:HorizontalBranchStars}

In the previous section, we found two 
GCs, NGC~1754 and NGC~2005, have SSP-derived
ages significantly (3$\sigma$) 
younger than the literature values.
\citeANP{Rabin82} (1982) was the first to suggest that
the presence of blue HB stars may have a significant effect
upon the equivalent width of the Balmer indices, possibly comparable
to the contribution from stars at the main sequence turn-off.
Since NGC~2005 has a largely blue HB, 
is it possible that the presence 
of blue HB stars are the origin of this present
disagreement with the SSP models?
Looking at this issue, \citeANP{deFreitasPacheco95} (1995) 
compared the H$\beta$ line-strengths of 10 Galactic GCs
with their HB morphologies.
With the aid of empirical modelling, these authors 
suggested that Blue HBs may increase H$\beta$ 
by upwards of $\sim$ 1.0 \AA.

For the five clusters in our sample with \HST\ CMDs
from \citeANP{Olsen98} (1998), we have an
accurate measure of their CMD age, 
HB morphology and integrated Balmer indices.
Since these clusters (NGC~1754, NGC~1835, NGC~1898, NGC~2005
and NGC~2019) all have similar mean spectroscopic
metallicities (mean metallicities derived from
Table~B1 are [Fe/H] = --1.38, --1.57, --1.30, --1.43
and --1.42 respectively), we are able to directly compare 
the effect of HB morphology upon
the Lick/IDS Balmer indices of the LMC clusters.

\begin{figure}
\centering
\centerline{\psfig{file=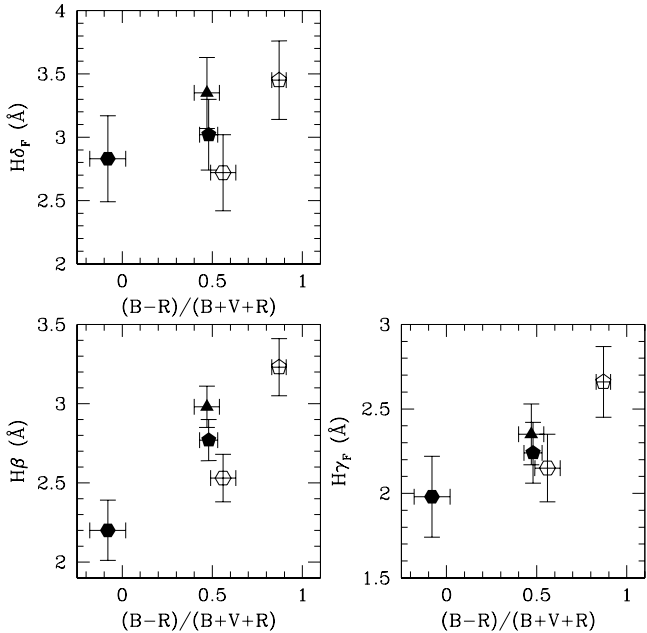,height=9cm}}
\caption{Balmer line-strength plotted against 
the HB parameter $(B-R)/(B+V+R)$ of five LMC
globular clusters.
Filled triangle : NGC~1754, filled pentagon : NGC~1835, 
filled hexagon : NGC~1898, open pentagon : NGC~2005 and
open hexagon : NGC~2019.}
\label{fig:hb}
\end{figure}

We plot the Lick/IDS Balmer indices of these GCs 
against their respective HB parameters 
in Figure~\ref{fig:hb}.
The HB parameter (\eg \citeANP{Lee94} 1994) 
is commonly given by $(B-R)/(B+V+R)$, where $B$ is the 
number of HB stars to the blue of the instability strip, 
$V$ the number of RR Lyrae stars and $R$ the number of 
HB stars redward of the instability strip.
Negative values of this parameter indicate a red
HB morphology (--1 corresponds to no 
blue HB stars), increasingly 
positive numbers correspond to progressively bluer 
HBs. 

As shown in Figure~\ref{fig:hb}, there is 
a correlation between HB parameter
and Balmer line-strength, in the sense that 
bluer HBs lead to larger index values
for H$\beta$, H$\gamma_{\rm F}$ and H$\delta_{\rm F}$.
NGC~1898, which has a roughly equal number of stars to the
left and right of its RR Lyrae variables (and has the reddest determined 
HB morphology in our sample), has H$\beta$ line-strengths
$\sim$ 1 \AA\ lower than that of NGC~2005, a cluster 
which has a blue HB. 

If we temporarily ignore the possible contribution
from HB stars, at a given metallicity, a change in $\sim$ 1 \AA\
in H$\beta$ would be interpreted as an age difference
of $>$ 10 Gyr using the (extrapolated) W94 models.
Since \citeANP{Olsen98} (1998) obtained a CMD age 
for NGC~1898 of 14.0 $\pm$ 2.3 Gyr, and for
NGC~2005 of 16.6 $\pm$ 5.1 Gyr (assuming 
\citeANP{Olszewski91} 1991 abundances), 
the possibility of a 10 Gyr age difference
between these two clusters seems unlikely.

One possible origin of this behaviour 
of the Balmer lines is the metallicity dependence
of the integrated indices themselves.
We found in this study that the metallicities 
of the clusters in Figure~\ref{fig:hb} are very similar.
This is also true of the metallicities 
derived by \citeANP{Olsen98} (1998) for these same clusters, 
using the $V^{TO}_{HB}$
method of \citeANP{Sarajedini94} (1994).
However, \citeANP{Olszewski91} (1991) found from their
Ca-Triplet observations that NGC~2005 was some $\sim$ --0.5 dex 
more metal-poor than NGC~1898. 
Since the Lick/IDS H$\beta$ index is really 
a blend of many metallic lines, centred on the
H$\beta$ feature, significant metallicity
differences amongst the clusters may be driving
variations in this index.

To test this possibility, we have used the W94 SSP
models which do not account for variations
in HB morphology, but treat HB stars purely as a red clump.
The predictions of these models are qualitatively very similar
to SSP models with no HB stars 
(\eg W94; \citeANP{Lee00a} 2000). 
The W94 models predict that H$\beta$ increases very slowly with 
decreasing metallicity at old ages. At 17 Gyr,
they indicate that H$\beta$ increases by
+0.12 \AA\ from [Fe/H] = --1.5 to [Fe/H] = --2.0.
Moreover, this increase in H$\beta$ becomes 
only slightly larger 
for younger ages, +0.17 \AA\ at 12 Gyr and
+ 0.27 \AA\ at 8 Gyr. 
Therefore, even if NGC~2005 is --0.5 dex more metal-poor
than NGC~1898, the metallicity sensitivity
of Balmer lines themselves cannot 
explain the position of the GCs in 
Figure~\ref{fig:hb}. 

We conclude that the Balmer-line indices
of the LMC GCs are significantly
effected (increased by up to $\sim$ 1.0 \AA\
in H$\beta$) by the presence of blue HB stars.
The \citeANP{Maraston00} (2000) and KFF99 models
try to account for this dependency of HB morphology on metallicity,
and its subsequent contribution to the Balmer lines, by 
including mass-loss of the RGB.
However, a shift of --0.5 dex in the \citeANP{Maraston00}
models corresponds to an increase in H$\beta$ 
of only $\sim$ 0.5 \AA\ at old ages and 
low metallicities, inconsistent
with the ages of NGC~1754 and NGC~2005.

This, however, is no real fault of the models; HB morphology
cannot not be directly predicted from first principles in stellar
evolution. The unknown mechanisms for mass-loss must be subsumed within an
analytical expression such as Reimer's formalism (among others) 
which has little physical basis. Moreover, the HB morphologies 
of GCs are not a monotonic function of metallicity, and require 
at least one more parameter to describe their morphology. 
This ``second parameter effect'' is thought to be age, although
the issue is far from settled (\eg \citeANP{Lee94} 1994).
Therefore, since SSP models cannot directly predict HB
morphologies, they must be calibrated from observations of real 
GCs (\eg \citeANP{Lee00a} 2000).

Finally, we note that in the \HST\ CMD for NGC~1898 from
\citeANP{Olsen98} (1998), there are a number of stars which
lie in the location on the CMD consistent with blue straggler
stars (\ie $V-I$ $\sim$ 0.2, $V \sim$ 21). 
Whilst \citeANP{Olsen98} (1998) make no such claim, due to 
uncertainties in their CMD-cleaning procedure, the possibility that 
they are blue stragglers is intriguing. The location
of this cluster on the SSP grids suggests that if these
are indeed blue straggler stars, they do not contribute 
significantly the Balmer indices of NGC~1898 (\eg see
\citeANP{Burstein84} 1984 and \citeANP{Trager00a} 2000b).

A more detailed comparative study of the blue straggler population, 
and their effects on integrated indices, should be 
performed between Galactic (or otherwise) GCs with very 
similar metallicities, ages and HB morphologies.

\section{Summary and Conclusions}
\label{sec:SummaryandConclusions}

We have obtained high S/N integrated spectra for 24 star clusters in the
LMC, and derived age and metallicity estimates for these
clusters using a combination of the Lick/IDS indices 
and the SSP models of \citeANP{Maraston00} (2000) 
and \citeANP{Kurth99} (1999).
To test the SSP models, we have compiled a list of metallicity
and age determinations from the literature.
Metallicities were taken largely from the Ca-Triplet 
spectroscopy of \citeANP{Olszewski91} (1991). 
The age estimates are somewhat more inhomogeneous
in nature, deriving from a number of methods; the 
location of the main-sequence turn-off in CMDs, 
the extent of the AGB calibration of \citeANP{Mould82} (1982)
or integrated colours. Comparing these independently
derived quantities we find:

\begin{itemize}

\item the SSP-derived metallicities, 
obtained using both the Mg$_2$ and $\langle$Fe$\rangle$ Lick/IDS 
indices show generally good agreement with the literature
values. However, the Mg$_2$ index (and to a lesser degree Mg $b$)
predicts metallicities which are systematically higher than
those from $\langle$Fe$\rangle$, by up to +0.5 dex for the
highest metallicity clusters. Amongst the possible explanations 
for this difference is the existence of [$\alpha$/Fe] $>$ 0 
in the clusters.
We publish metallicities for 11 LMC star clusters with no
previous measurements, which are accurate to $\sim$ 0.2 dex.

\item for the majority of the LMC clusters, the SSP-models predict ages 
from the H$\beta$, H$\gamma_{\rm F}$ and H$\delta_{\rm F}$
indices which are consistent with the literature values.
However, age estimates of the old
LMC globular clusters are often ambiguous.
The oldest isochrones of the SSP models 
overlap younger isochrones due to the 
modelling of mass-loss on the RGB.
{\it Assuming} old ages in interpreting these
data, six clusters, NGC~1786, NGC~1835, 
NGC~1898, NGC~1916, NGC~1939 and NGC~2019, 
have SSP-derived ages in all three
measurable Balmer indices 
which are consistent with their ages
derived from CMDS or integrated colours.

\item two clusters, namely NGC~1754 and NGC~2005, 
have extremely strong Balmer lines, which leads 
to the SSP model ages which are too young ($\sim$ 8 and 6 Gyr
respectively).
Comparison between the horizontal branch morphology
and the Balmer lines for five of the GCs in our sample
suggests that blue HBs are likely contributing 
up to $\sim$ 1.0 \AA\ to the H$\beta$ index in these clusters.

\end{itemize}

We conclude that the SSP models considered
in this study are able to satisfactorily predict
the ages and metallicities for the vast majority
of LMC star clusters from integrated
spectroscopic indices. This remains true despite
the rather inhomogenous nature of the literature
age determinations.
However, estimating the ages of the old, low-metallicity
LMC GCs is severely complicated by the strong contribution
of horizontal branch/post-horizontal branch stars to integrated
indices. We conclude that at old ages and low metallicities 
([Fe/H] $<$ --1.0), Balmer lines (\eg H$\beta$, H$\gamma$, H$\delta$)
are not useful age indicators without {\it a priori} knowledge of
horizontal branch morphology.

\section{Acknowledgements}

We wish to thank Hyun-chul Lee for useful 
comments regarding this paper, Mike Reid for the preparation of 
UKST finding charts, and Malcolm Hartley, 
who went above and beyond the call 
of duty during the FLAIR observations. 
We are in debt to Claudia Maraston
for interesting discussions and for providing her full 
models ahead of publication, and the anonymous referee
for their detailed appraisal of the paper.
MB acknowledges the Royal Society for its 
fellowship grant.

\appendix

\section{Corrected Lick Indices for LMC Star Clusters}
\label{app:CorrectedLickIndicesforLMCStarClusters}

The Lick/IDS indices for 24 LMC star clusters after the additive
corrections are applied (Table~\ref{tab:offsets})
are given in Table~A1.
The uncertainties are tabulated in alternate rows, 
calculated from the S/N of the spectra added in quadrature
to our index repeatability which is given in Table~\ref{tab:repeats}.

\begin{table*}
\caption{Lick/IDS indices for LMC star clusters. Uncertainties
tabulated in alternate rows are calculated from the S/N of the
spectra added in quadrature to our repeatability of each index.}
\label{tab:indices}
\begin{tabular}{lrrrrrrrrrr} 
\hline
ID & CN$_1$ & CN$_2$ & Ca4227 & G4300 & Fe4383 & Ca4455 &Fe4531&C4668 & H$\beta$ & Fe5015\\
   & (mag) & (mag) & (\AA) & (\AA) & (\AA) & (\AA) & (\AA) &(\AA) & (\AA)& (\AA)\\
\hline
NGC~1718   & --0.141  & --0.087  & 0.267  & 1.747  & 0.549  & 0.167  & 0.966  & --0.183  & 3.265  & 2.444\\
   $\pm$   & 0.018  & 0.021  & 0.153  & 0.310  & 0.366  & 0.196  & 0.966  & 0.413  & 0.169  & 0.428\\
NGC~1751   & --0.142  & --0.112  & 1.158  & 2.360  & 1.978  & 0.255  & 1.565  & 2.083  & 4.122  & 4.396\\
   $\pm$    & 0.019  & 0.023  & 0.158  & 0.330  & 0.401  & 0.219  & 1.565  & 0.451  & 0.177  & 0.449\\
NGC~1754   & --0.122  & --0.082  & 0.830  & 1.707  & 0.909  & 0.156  & 1.349  & --0.218  & 2.981  & 2.755\\
   $\pm$   & 0.011  & 0.013  & 0.100  & 0.238  & 0.235  & 0.135  & 1.349  & 0.286  & 0.127  & 0.349\\
NGC~1786   & --0.084  & --0.040  & 0.468  & 1.771  & 0.743  & 0.450  & 1.273  & --0.666  & 2.210  & 1.662\\
   $\pm$   & 0.006  & 0.008  & 0.075  & 0.205  & 0.164  & 0.106  & 1.273  & 0.219  & 0.110  & 0.317\\
NGC~1801   & --0.188  & --0.127  & 0.237  & --2.320  & --0.607  & 0.281  & 0.675  & --0.662  & 5.694  & 1.737\\
   $\pm$   & 0.009  & 0.012  & 0.092  & 0.235  & 0.226  & 0.132  & 0.675  & 0.290  & 0.129  & 0.362\\
NGC~1806   & --0.099  & --0.041  & 0.432  & 1.184  & 0.642  & 0.866  & 2.175  & 0.784  & 3.230  & 3.616\\
   $\pm$  & 0.011  & 0.013  & 0.103  & 0.241  & 0.240  & 0.137  & 2.175  & 0.288  & 0.131  & 0.354\\
NGC~1830   & --0.207  & --0.141  & 0.600  & --1.911  & --1.093  & --0.503  & 1.692  & 0.319  & 4.953  & 2.051\\
   $\pm$  & 0.020  & 0.025  & 0.173  & 0.374  & 0.451  & 0.238  & 1.692  & 0.507  & 0.197  & 0.506\\
NGC~1835   & --0.106  & --0.054  & 0.403  & 1.662  & 0.832  & 0.426  & 1.320  & --0.237  & 2.771  & 2.531\\
   $\pm$   & 0.010  & 0.013  & 0.098  & 0.233  & 0.225  & 0.132  & 1.320  & 0.274  & 0.125  & 0.343\\
NGC~1846   & --0.176  & --0.118  & 0.880  & 0.882  & 0.856  & 0.470  & 2.498  & --0.265  & 3.672  & 2.752\\
   $\pm$   & 0.014  & 0.016  & 0.120  & 0.269  & 0.283  & 0.156  & 2.498  & 0.326  & 0.139  & 0.373\\
NGC~1852   & --0.122  & --0.069  & 0.769  & 1.033  & 0.290  & 0.273  & 1.786  & 0.201  & 3.723  & 3.436\\
   $\pm$   & 0.013  & 0.015  & 0.117  & 0.265  & 0.288  & 0.160  & 1.786  & 0.340  & 0.146  & 0.385\\
NGC~1856   & --0.205  & --0.129  & 0.538  & --1.447  & 0.389  & 0.334  & 1.222  & 0.040  & 6.076  & 2.568\\
   $\pm$   & 0.007  & 0.009  & 0.077  & 0.212  & 0.175  & 0.111  & 1.222  & 0.232  & 0.112  & 0.325\\
NGC~1865   & --0.224  & --0.174  & 0.566  & --0.473  & 0.164  & 0.301  & 1.295  & --0.646  & 5.506  & 1.844\\
   $\pm$   & 0.015  & 0.018  & 0.133  & 0.297  & 0.343  & 0.186  & 1.295  & 0.409  & 0.164  & 0.441\\
NGC~1872   & --0.201  & --0.139  & 0.255  & --1.685  & --0.042  & 0.492  & 0.919  & --0.609  & 5.269  & 0.745\\
   $\pm$   & 0.014  & 0.016  & 0.124  & 0.285  & 0.311  & 0.170  & 0.919  & 0.379  & 0.155  & 0.422\\
NGC~1878   & --0.218  & --0.141  & 0.481  & --1.021  & --0.682  & 0.237  & 2.061  & --0.427  & 5.824  & 2.371\\
   $\pm$   & 0.010  & 0.012  & 0.093  & 0.234  & 0.221  & 0.129  & 2.061  & 0.273  & 0.122  & 0.346\\
NGC~1898   & --0.074  & --0.017  & 0.646  & 2.634  & 1.044  & 1.452  & 1.910  & 0.583  & 2.191  & 2.195\\
   $\pm$    & 0.021  & 0.025  & 0.178  & 0.346  & 0.439  & 0.223  & 1.910  & 0.479  & 0.190  & 0.463\\
NGC~1916   & --0.117  & --0.079  & 0.196  & 0.889  & 0.882  & 0.334  & 1.560  & --0.001  & 2.840  & 2.075\\
   $\pm$   & 0.011  & 0.013  & 0.102  & 0.240  & 0.239  & 0.138  & 1.560  & 0.293  & 0.132  & 0.355\\
NGC~1939   & --0.102  & --0.051  & 0.553  & 0.729  & --0.168  & 0.268  & 0.325  & --0.605  & 2.628  & 2.056\\
   $\pm$   & 0.013  & 0.015  & 0.117  & 0.260  & 0.285  & 0.159  & 0.325  & 0.340  & 0.144  & 0.384\\
NGC~1978   & --0.106  & --0.070  & 0.820  & 2.746  & 1.766  & 0.702  & 2.465  & 2.597  & 3.188  & 3.986\\
   $\pm$   & 0.012  & 0.014  & 0.106  & 0.243  & 0.242  & 0.138  & 2.465  & 0.280  & 0.127  & 0.346\\
NGC~1987   & --0.204  & --0.154  & 0.422  & 0.151  & 1.055  & 0.494  & 2.054  & 1.424  & 3.911  & 3.675\\
   $\pm$   & 0.012  & 0.014  & 0.110  & 0.257  & 0.262  & 0.149  & 2.054  & 0.316  & 0.139  & 0.370\\
NGC~2005   & --0.113  & --0.047  & 0.533  & 0.934  & 1.513  & 0.669  & 1.327  & --2.303  & 3.227  & 2.150\\
   $\pm$    & 0.016  & 0.020  & 0.150  & 0.312  & 0.364  & 0.198  & 1.327  & 0.449  & 0.175  & 0.445\\
NGC~2019   & --0.103  & --0.049  & 0.762  & 0.892  & 0.527  & 0.328  & 1.960  & 0.253  & 2.531  & 2.005\\
   $\pm$    & 0.015  & 0.017  & 0.126  & 0.276  & 0.307  & 0.169  & 1.960  & 0.364  & 0.153  & 0.398\\
NGC~2107   & --0.200  & --0.143  & 0.592  & --1.340  & 0.003  & 0.475  & 1.503  & --1.499  & 5.649  & 1.649\\
   $\pm$    & 0.010  & 0.012  & 0.091  & 0.234  & 0.219  & 0.129  & 1.503  & 0.277  & 0.123  & 0.348\\
NGC~2108   & --0.164  & --0.120  & 1.190  & 1.239  & --0.021  & --0.082  & 2.361  & 0.565  & 3.828  & 2.207\\
   $\pm$   & 0.018  & 0.021  & 0.144  & 0.312  & 0.361  & 0.192  & 2.361  & 0.392  & 0.160  & 0.411\\
SL~250    & --0.148  & --0.081  & 0.506  & --0.465  & --0.960  & 0.189  & 1.488  & --0.095  & 4.223  & 3.528\\
   $\pm$   & 0.012  & 0.014  & 0.106  & 0.253  & 0.266  & 0.148  & 1.488  & 0.316  & 0.139  & 0.374\\
\hline
\end{tabular}
\end{table*}

\begin{table*}
\contcaption{}
\begin{tabular}{lrrrrrrrrrr} 
\hline
ID & Mg$_1$ & Mg$_2$ & Mg $b$ & Fe5270 &Fe5335 & Fe5406 &H$\delta_{\rm A}$ & H$\gamma_{\rm A}$ & H$\delta_{\rm F}$ & H$\gamma_{\rm F}$\\
	&(mag) & (mag)& (\AA) & (\AA) & (\AA) & (\AA) & (\AA) &(\AA) & (\AA)& (\AA)\\
\hline
NGC~1718   & 0.029  & 0.090  & 1.316  & 1.057  & 1.027  & 1.179  & 6.168  & 3.321  & 4.761  & 4.038\\
  $\pm$    & 0.007  & 0.009  & 0.164  & 0.199  & 0.213  & 0.183  & 0.407  & 0.279  & 0.317  & 0.208\\
NGC~1751   & 0.036  & 0.112  & 1.815  & 1.634  & 1.261  & 0.697  & 6.288  & 3.245  & 4.880  & 4.478\\
  $\pm$    & 0.008  & 0.009  & 0.177  & 0.211  & 0.234  & 0.199  & 0.418  & 0.303  & 0.324  & 0.217\\
NGC~1754   & 0.022  & 0.071  & 1.451  & 1.164  & 0.886  & 0.430  & 4.462  & 1.825  & 3.354  & 2.345\\
  $\pm$    & 0.004  & 0.006  & 0.109  & 0.146  & 0.150  & 0.140  & 0.347  & 0.214  & 0.282  & 0.180\\
NGC~1786   & 0.020  & 0.062  & 1.274  & 0.753  & 0.703  & 0.160  & 3.065  & 1.200  & 2.251  & 2.172\\
  $\pm$    & 0.002  & 0.004  & 0.085  & 0.123  & 0.118  & 0.120  & 0.320  & 0.185  & 0.266  & 0.167\\
NGC~1801   & 0.014  & 0.044  & 0.690  & 0.530  & 0.429  & --0.335  & 9.074  & 8.370  & 6.372  & 6.076\\
  $\pm$    & 0.005  & 0.007  & 0.123  & 0.161  & 0.171  & 0.156  & 0.330  & 0.199  & 0.271  & 0.172\\
NGC~1806   & 0.051  & 0.112  & 1.668  & 1.659  & 1.063  & 1.196  & 5.202  & 3.547  & 3.582  & 3.190\\
  $\pm$    & 0.004  & 0.006  & 0.116  & 0.149  & 0.155  & 0.143  & 0.346  & 0.215  & 0.282  & 0.180\\
NGC~1830   & --0.001  & 0.062  & 1.192  & 0.843  & 0.641  & --0.035  & 8.152  & 7.026  & 5.380  & 5.183\\
  $\pm$    & 0.009  & 0.011  & 0.208  & 0.248  & 0.282  & 0.235  & 0.422  & 0.298  & 0.331  & 0.221\\
NGC~1835   & 0.024  & 0.074  & 1.359  & 0.878  & 0.755  & 0.251  & 4.022  & 1.361  & 3.019  & 2.238\\
  $\pm$   & 0.004  & 0.006  & 0.106  & 0.142  & 0.145  & 0.136  & 0.343  & 0.211  & 0.279  & 0.178\\
NGC~1846   & 0.025  & 0.100  & 1.617  & 1.590  & 1.101  & 0.605  & 6.271  & 3.456  & 4.450  & 3.219\\
  $\pm$    & 0.005  & 0.007  & 0.128  & 0.165  & 0.173  & 0.156  & 0.372  & 0.237  & 0.298  & 0.190\\
NGC~1852   & 0.033  & 0.098  & 1.576  & 1.384  & 1.107  & 0.610  & 6.263  & 3.285  & 4.427  & 3.237\\
  $\pm$    & 0.006  & 0.008  & 0.137  & 0.173  & 0.185  & 0.163  & 0.356  & 0.236  & 0.288  & 0.190\\
NGC~1856   & 0.022  & 0.074  & 1.428  & 0.870  & 1.141  & 0.579  & 9.626  & 8.066  & 6.894  & 6.403\\
  $\pm$    & 0.003  & 0.004  & 0.093  & 0.130  & 0.129  & 0.127  & 0.319  & 0.185  & 0.265  & 0.166\\
NGC~1865   & 0.032  & 0.094  & 1.767  & 1.031  & 0.912  & --0.154  & 10.400  & 7.051  & 6.511  & 5.707\\
  $\pm$    & 0.008  & 0.009  & 0.171  & 0.213  & 0.234  & 0.201  & 0.363  & 0.250  & 0.292  & 0.195\\
NGC~1872   & 0.010  & 0.065  & 1.423  & 0.729  & 0.207  & 0.224  & 9.446  & 8.169  & 6.471  & 5.926\\
  $\pm$    & 0.007  & 0.009  & 0.159  & 0.199  & 0.223  & 0.192  & 0.357  & 0.234  & 0.288  & 0.188\\
NGC~1878   & 0.019  & 0.072  & 1.237  & 0.737  & 1.191  & 0.625  & 10.980  & 8.177  & 7.518  & 6.183\\
  $\pm$    & 0.004  & 0.006  & 0.109  & 0.145  & 0.148  & 0.139  & 0.330  & 0.199  & 0.272  & 0.172\\
NGC~1898   & 0.045  & 0.095  & 1.951  & 1.057  & 1.419  & 0.842  & 1.028  & --0.071  & 2.830  & 1.980\\
  $\pm$    & 0.008  & 0.009  & 0.176  & 0.212  & 0.228  & 0.194  & 0.459  & 0.333  & 0.341  & 0.236\\
NGC~1916   & 0.016  & 0.056  & 1.032  & 0.765  & 0.703  & 0.187  & 4.841  & 2.565  & 3.339  & 2.635\\
  $\pm$    & 0.004  & 0.006  & 0.115  & 0.150  & 0.156  & 0.144  & 0.342  & 0.215  & 0.279  & 0.180\\
NGC~1939   & 0.012  & 0.065  & 1.560  & 0.638  & 0.694  & 0.593  & 4.220  & 3.119  & 3.017  & 2.723\\
  $\pm$    & 0.006  & 0.007  & 0.134  & 0.170  & 0.181  & 0.159  & 0.359  & 0.234  & 0.289  & 0.189\\
NGC~1978   & 0.056  & 0.141  & 1.938  & 1.772  & 1.539  & 0.573  & 3.493  & 0.755  & 2.566  & 1.888\\
  $\pm$    & 0.004  & 0.006  & 0.109  & 0.143  & 0.147  & 0.138  & 0.359  & 0.223  & 0.289  & 0.184\\
NGC~1987   & 0.041  & 0.121  & 1.775  & 1.341  & 0.960  & 0.622  & 8.526  & 5.645  & 5.453  & 4.553\\
  $\pm$    & 0.005  & 0.007  & 0.127  & 0.164  & 0.175  & 0.156  & 0.347  & 0.222  & 0.283  & 0.183\\
NGC~2005   & 0.013  & 0.065  & 1.343  & 0.803  & 1.127  & 0.257  & 4.613  & 2.246  & 3.446  & 2.661\\
  $\pm$    & 0.008  & 0.009  & 0.171  & 0.209  & 0.229  & 0.197  & 0.392  & 0.282  & 0.310  & 0.212\\
NGC~2019   & 0.012  & 0.075  & 1.201  & 1.175  & 0.730  & 0.686  & 3.305  & 2.257  & 2.725  & 2.157\\
  $\pm$    & 0.006  & 0.008  & 0.143  & 0.178  & 0.194  & 0.169  & 0.374  & 0.248  & 0.298  & 0.197\\
NGC~2107   & 0.022  & 0.064  & 1.067  & 1.005  & 1.086  & 0.198  & 9.695  & 7.542  & 6.800  & 5.885\\
  $\pm$    & 0.004  & 0.006  & 0.111  & 0.147  & 0.152  & 0.143  & 0.332  & 0.199  & 0.273  & 0.172\\
NGC~2108   & 0.040  & 0.108  & 1.385  & 1.599  & 1.760  & 0.916  & 7.052  & 6.401  & 4.768  & 4.988\\
  $\pm$    & 0.007  & 0.009  & 0.151  & 0.184  & 0.193  & 0.171  & 0.411  & 0.266  & 0.323  & 0.204\\
SL~250    & 0.018  & 0.085  & 1.292  & 1.242  & 0.812  & 0.232  & 7.878  & 6.355  & 5.358  & 4.536\\
  $\pm$    & 0.005  & 0.007  & 0.132  & 0.168  & 0.181  & 0.161  & 0.345  & 0.217  & 0.281  & 0.181\\
\hline
\end{tabular}
\end{table*}

\section{Age and Metallicity Predictions of SSP Models}
\label{app:AgeandMetallicityPredictionsofSSPModels}

Table~B1 lists the age and metallicities of the LMC clusters
derived using the Maraston SSP models, from the
H$\beta$--Mg$_2$ and H$\gamma_{\rm F}$--$\langle$Fe$\rangle$
Lick/IDS indices. Also tabulated are the available literature 
ages and metallicities for these clusters.

\begin{table*}
\caption{Ages and metallicities of LMC star clusters.}
\label{tab:results}
\begin{tabular}{lcrcrccc} 
\hline
ID & [Fe/H] & Age (Gyr) & [Fe/H]  & Age (Gyr) & [Fe/H] & Age (Gyr) & Sources\\
& (Mg$_2$,H$\beta$) & (Mg$_2$,H$\beta$) &
($\langle$Fe$\rangle$,H$\gamma_{\rm F}) $ &
($\langle$Fe$\rangle$,H$\gamma_{\rm F}$) & (literature) & (literature) &  \\
\hline
NGC~1718 & --1.12$_{-0.22}^{+0.18}$ & 4.91$_{-0.74}^{+0.53}$ & --0.98$_{-0.30}^{+0.29}$ & 1.98$_{-0.53}^{+0.88}$ & ... & 1.81$\pm$0.31&...,3\\
NGC~1751 & --0.39$_{-0.32}^{+0.25}$ & 2.31$_{-0.72}^{+0.43}$ & --0.42$_{-0.23}^{+0.29}$ & 0.92$_{-0.06}^{+0.25}$ & --0.18$\pm$0.20 & 1.48$\pm$0.55&5,1\\
NGC~1754 & --1.37$_{-0.12}^{+0.21}$ & 7.00$_{-2.00}^{+1.50}$ &--1.38$_{-0.14}^{+0.13}$ & 14.00$_{-0.54}^{+1.00}$ &--1.54$\pm$0.20(--1.42$\pm$0.15*) & 15.6$\pm$2.3(15.6$\pm$2.2*)&5,8\\
NGC~1786 & --1.58$_{-0.12}^{+0.13}$ & 15.30$_{-1.00}^{+1.40}$&--1.63$_{-0.12}^{+0.11}$ & ... &--1.87$\pm$0.20 & 15.1$\pm$3.1 & 5,7\\
NGC~1801 & --1.01$_{-0.32}^{+0.23}$ & 0.47$_{-0.04}^{+0.02}$ & --0.98$_{-0.28}^{+0.27}$ & 0.80$_{-0.57}^{+0.28}$ & ... & 0.30$\pm$0.10&...,11\\
NGC~1806 & --0.71$_{-0.16}^{+0.15}$ & 3.49$_{-0.47}^{+0.05}$ & --0.73$_{-0.16}^{+0.17}$ & 1.59$_{-0.39}^{+0.53}$ & --0.71$\pm$0.74& 0.50$\pm$0.10&10\\
NGC~1830 & --1.02$_{-0.40}^{+0.19}$ & 1.23$_{-0.89}^{+0.38}$ & --1.30$_{-0.17}^{+0.51}$ & 1.50$_{-0.29}^{+0.63}$ & ... & 0.30$\pm$0.10&...,11\\
NGC~1835 & --1.40$_{-0.13}^{+0.18}$ & 8.31$_{-1.80}^{+1.80}$ &--1.74$_{-0.16}^{+0.22}$ &12.50$_{-2.00}^{+1.50}$&--1.72$\pm$0.20(--1.62$\pm$0.15*)&16.6$\pm$2.9(16.2$\pm$2.8*)&5,8\\
NGC~1846 & --0.80$_{-0.14}^{+0.14}$ & 3.10$_{-0.46}^{+0.07}$ & --0.75$_{-0.18}^{+0.20}$ & 1.69$_{-0.61}^{+0.67}$&--0.70$\pm$0.20&2.85$\pm$1.10&5,1\\
NGC~1852 & --0.85$_{-0.15}^{+0.15}$ & 2.99$_{-0.36}^{+0.08}$ & --0.88$_{-0.19}^{+0.20}$ & 2.30$_{-0.36}^{+1.03}$&...&2.51$\pm$0.93&...,1\\
NGC~1856 & --0.25$_{-0.18}^{+0.19}$ & 0.60$_{-0.10}^{+0.06}$ & --0.09$_{-0.10}^{+0.19}$ & 0.34$_{-0.09}^{+0.02}$&...&0.12$\pm$0.03&...,4\\
NGC~1865 & --0.41$_{-0.12}^{+0.33}$ & 0.72$_{-0.05}^{+0.07}$ & --0.46$_{-0.27}^{+0.25}$ & 0.63$_{-0.21}^{+0.12}$&...&0.89$\pm$0.33&...,7\\
NGC~1872 & --0.72$_{-0.30}^{+0.10}$ & 1.02$_{-0.21}^{+0.12}$ & --0.72$_{-0.16}^{+0.12}$ & 0.14$_{-0.18}^{+0.21}$&...&0.30$\pm$0.10&...,11\\
NGC~1878 & --0.29$_{-0.35}^{+0.15}$ & 0.37$_{-0.15}^{+0.18}$ & --0.24$_{-0.14}^{+0.14}$ & 0.38$_{-0.02}^{+0.02}$&...&0.30$\pm$0.10&...,11\\
NGC~1898 & --1.27$_{-0.15}^{+0.20}$ & 11.01$_{-1.83}^{+2.00}$ & --1.32$_{-0.15}^{+0.33}$ & 13.70$_{-1.00}^{+1.30}$&--1.37$\pm$0.20(--1.18$\pm$0.16*)&14.0$\pm$2.3(13.5$\pm$2.2*)&5,8\\
NGC~1916 & --2.10$_{-0.19}^{+0.20}$ & 15.30$_{-2.23}^{+4.00}$  &--1.80$_{-0.20}^{+0.20}$ & 14.54$_{-1.50}^{+3.00}$ &--2.08$\pm$0.20&10.5$\pm$5.5&5,11\\ 
NGC~1939 & --1.55$_{-0.36}^{+0.10}$ & 15.00$_{-2.03}^{+3.02}$ & --2.01$_{-0.23}^{+0.36}$ & 15.20$_{-2.00}^{+4.10}$&--2.00$\pm$0.20&10.5$\pm$5.5&9,11\\
NGC~1978 & --0.21$_{-0.26}^{+0.25}$ & 1.52$_{-0.38}^{+0.32}$ & --0.58$_{-0.16}^{+0.15}$ & 1.34$_{-0.94}^{+2.41}$&--0.42$\pm$0.20&2.00$\pm$0.74&5,7\\
NGC~1987 & --0.42$_{-0.14}^{+0.26}$ & 2.42$_{-0.06}^{+0.31}$ &--0.79$_{-0.19}^{+0.21}$ & 0.84$_{-0.22}^{+0.03}$&--0.50$\pm$0.20&2.51$\pm$0.93&2,1 \\
NGC~2005 & --1.51$_{-0.31}^{+0.12}$ & 6.27$_{-1.96}^{+1.22}$ & --1.34$_{-0.32}^{+0.26}$ & 16.00$_{-1.50}^{+1.99}$&--1.92$\pm$0.20(--1.35$\pm$0.15*)&16.6$\pm$5.1(15.5$\pm$4.9*)&5,8\\
NGC~2019 & --1.41$_{-0.20}^{+0.40}$ & 16.00$_{-3.04}^{+2.22}$ & --1.44$_{-0.37}^{+0.16}$ & 13.30$_{-1.00}^{+0.80}$&--1.81$\pm$0.20(--1.23$\pm$0.15*)&17.8$\pm$3.2(16.3$\pm$3.1*)&5,8\\
NGC~2107 & --0.64$_{-0.56}^{+0.21}$ & 0.46$_{-0.35}^{+0.23}$ & --0.30$_{-0.14}^{+0.13}$ & 0.48$_{-0.04}^{+0.02}$&...&1.00$\pm$0.37&...,1\\
NGC~2108 & --0.74$_{-0.16}^{+0.35}$ & 3.02$_{-0.10}^{+0.46}$ & --0.22$_{-0.22}^{+0.22}$ & 0.56$_{-0.09}^{+0.03}$&...&0.59$\pm$0.22&...,6\\
SL~250 & --1.00$_{-0.12}^{+0.37}$ & 2.86$_{-0.40}^{+0.42}$ & --0.97$_{-0.18}^{+0.21}$ & 1.24$_{-0.28}^{+0.46}$&...&0.60$\pm$0.20&...,11\\
\hline
\end{tabular}
Sources: 1 = CMD, \citeANP{Mould82} (1982); 2 = Integrated
Spectroscopy, \citeANP{Rabin82} (1982); 3 = CMD,
\citeANP{ElsonandFall88} (1988); 
4 = CMD, \citeANP{Hodge84} (1984); 5 = Spectroscopy, \citeANP{Olszewski91} (1991); 
6 = CMD, \citeANP{Corsi94} (1994); 7 = CMD, \citeANP{Geisler97} (1997);
8 = \HST\ CMD, \citeANP{Olsen98} (1998); 9 = Integrated
Spectroscopy, \citeANP{Dutra99} (1999); 10 = Str\"{o}mgren
photometry, \citeANP{Dirsch00} (2000);\newline 11 = ages inferred
from the SWB type of the cluster.  * metallicities obtained using the method of
\citeANP{Sarajedini94} (1994).
\end{table*}

\bibliographystyle{mnras}
\bibliography{mnras}    
     
\end{document}